\documentclass[10pt,final,doublecolumn]{IEEEtran}
\IEEEoverridecommandlockouts
\usepackage{amsmath}
\usepackage{amssymb}
\usepackage{amsthm}
\usepackage{mathrsfs}
\usepackage{latexsym}
\usepackage{graphicx}
\usepackage{bbding}
\usepackage{indentfirst}
\usepackage{cases}
\usepackage{supertabular}
\usepackage{algorithm,algorithmic}
\usepackage{subeqnarray}
\usepackage{color}
\usepackage{bm}
\usepackage{stfloats}

\usepackage{subfigure}
\usepackage{array}
\usepackage{stfloats}
\usepackage{algorithmic}
\usepackage{float}
\usepackage{algorithm}
\allowdisplaybreaks[4]

\begin{document}
\title{Joint Communication and Sensing Design for Integrated Satellite-Terrestrial Maritime Systems}
\author{\IEEEauthorblockN{
Kaiwei Xiong, Xiaoming Chen, and Ming Ying}
\thanks{Kaiwei Xiong, Xiaoming Chen, and Ming Ying are with the College of Information Science and Electronic Engineering, Zhejiang University, Hangzhou 310027, China (e-mail:\{xiong\_kaiwei, chen\_xiaoming, and ming\_ying\}@zju.edu.cn).}
}\maketitle

\begin{abstract}
Joint communication and sensing has been a key technology in 6G. By integrating sensing into maritime communications, ships can communicate with the base station while sensing the surrounding environment to ensure safe navigation. In this paper, we introduce an integrated satellite-terrestrial maritime system (ISTMS) with joint communication and sensing based on the same radio-frequency signals. Specifically, the terrestrial base station (TBS) and low Earth orbit (LEO) satellite provide communication services for near-shore users (NSUs) and off-shore users (OSUs), respectively, while simultaneously performing target sensing. Based on a differential evolution method (DE), we propose a sensing algorithm, which can enhance the location accuracy and reduce resource consumption. Furthermore, we derive the key performance metrics for both communication and sensing. Through joint beamforming optimization of the TBS and LEO satellite, we maximize the sum rate of maritime users while satisfying target localization accuracy requirements and transmit power constraints. Finally, extensive simulation results demonstrate the effectiveness of the proposed algorithms in terms of location accuracy and transmission rate compared with the baseline algorithms.
	
\end{abstract}


\section{Introduction}
Recently, with the rapid development of the marine economy, maritime activities such as shipping, fishery, energy and environmental monitoring, have increased rapidly \cite{Maritime transport1}, \cite{Maritime transport2}. More than $80\%$ of global trade volume relies on maritime transportation. Trends such as containerization, giant ships, cargo tracking, and port automation have put forward unprecedented requirements for highly reliable, low-latency, and long-distance ship-to-shore communication. Due to the continuous increase of global aquaculture volume, maritime environment monitoring are the urgent needs. To achieve high-quality maritime communication, it is imperative to implement real-time monitoring of vessel positions, operational trajectories, and status parameters, coupled with instantaneous transmission rate to facilitate comprehensive maritime supervision. These capabilities constitute fundamental requirements for advanced maritime communication systems \cite{growing demand1}, \cite{growing demand2}.

At present, maritime communication systems primarily comprise two distinct components according to recent studies \cite{st1}: the coastal maritime base station network serving near-shore regions and satellite communication system providing services for off-shore areas. Unlike terrestrial communication networks, the deployment of maritime base stations faces unique environmental constraints. Traditional terrestrial base stations (TBS) offer severely limited maritime coverage, proving inadequate for off-shore communication demands. Moreover, the spectrum resources at sea are strictly divided, and the independent operation of each system leads to a low spectrum efficiency. The available maritime spectrum resources are strictly segmented and allocated, consequently degrading spectrum efficiency owing to the isolated operation of disparate communication systems. This fragmentation leads to substantial underutilization of valuable spectral resources that could otherwise enhance maritime connectivity. In addition, the means for sensing and monitoring maritime targets are scarce and costly. Some global navigation satellite systems (GNSS) can offer primary navigation with wide coverage. However, the GNSS is vulnerable to interference at sea, and their localization accuracy declines in the complex maritime environments.

To deal with the problems of insufficient coverage, low spectrum efficiency and high delay in the traditional maritime communication, the integrated satellite-terrestrial maritime system (ISTMS) has been proposed \cite{st2}, \cite{st3}, which combines the advantages of high throughput of TBS and wide coverage of low Earth orbit (LEO) satellites. Traditional maritime sensing mainly rely on shore-based and ship-borne dedicated radar. It causes high deployment costs, limited coverage, difficult information fusion, and is also independent of communication networks. In recent years, the application of LEO satellite constellations in localization and navigation has attracted widespread attention \cite{wq1}, \cite{wq2}. LEO satellites are deployed at altitudes ranging from 200 to 2,000 km, which has the advantages of low launch cost and high flexibility. In the current maritime communication network, communication and sensing systems are often independent, which leads to repetitive infrastructure construction, competition for spectrum resources, information silos and huge energy consumption. Recently, many studies focus on the integrated sensing and communication (ISAC) in terrestrial communication \cite{ISAC1}-\cite{ISAC4}, which can simultaneously support both communication and sensing functions and achieve in-depth resource sharing and optimization.

Many research scholars have already believed that ISAC will become a key technology in the future wireless systems, which can support lots of important application scenarios \cite{cj1}-\cite{cj3}. In \cite{tl1}, it proposed an integrated communication and radar beamforming framework, aiming to maximize the overall spectral efficiency in a multi-user and multi-target ISAC framework under the imperfect channel state information (CSI). The authors in \cite{tl2} investigated the ISAC dual in a reconfigurable intelligent surface (RIS)-aided communication system. Moreover, in the proposed system framework, the authors improved the spectral efficiency by fusing the sensing and communication signals that occupy the same frequency resources in the superimposed symbol. A safe ISAC system is introduced by \cite{tl3}, where distributed access points work together to communicate with users and sense targets in the presence of multiple eavesdroppers. It also design an effective joint optimization algorithm for communication and sensing beamforming vectors. In \cite{tl4}, it proposed an ISAC network based on the ship-based station, which can simultaneously provide maritime communication services and conduct target sensing. It also established a channel model to capture the complexity of the maritime environment and analyze the impact of the dynamic behavior of ship-based station on the ISAC network.

Although, the research schemes on ISAC mentioned above often focus on enhancing the overall performance of the network or the general sensing performance, while neglecting the refined information extraction and utilization of specific sensing targets. In \cite{dw1}, the authors proposed an ISAC framework based on RIS that can transmit and reflect simultaneously, dividing the entire space into a sensing space and a communication space. They also derived the Cramer-Rao bound (CRB) for two-dimensional wave arrival direction estimation for the localization of target. Based on the propagation characteristics of the signal, the authors derived the localization performance boundary and focused on maximizing the communication capacity and localization accuracy through a hybrid analog-digital precoding design \cite{dw2}. In addition to using the ISAC network to locate the position of the target \cite{dw3}, \cite{dw4}, it is also possible to perceive some other information such as the reflection coefficient, shapes, status and speed of the target. In \cite{dw3}, the authors proposed a new adaptive environmental sensing algorithm, which combines the two processes of multi-user information detection and environmental target detection, achieving precise environmental sensing. Besides, \cite{gz2} proposed an ISAC scheme based on multi-beams, in which some beams used for communication and other beams for sensing the environmental information.

However, the studies mentioned above mostly focus on single base station or terrestrial base stations, and do not analyze collaboration among different systems. Moreover, most of the recent researches on ISAC remains at the terrestrial communication systems and there is still a lack of research and analysis on maritime communication systems. Driven by these, our goal is to exploit the potential of joint communication and sensing design for the maritime communication systems. Thus, we propose a joint communication and sensing design for ISTMS. {Unlike prior ISAC works that focus on isolated nodes, this paper proposes a collaborative architecture to coordinate TBS and LEO satellites, while integrating maritime-specific physical constraints, such as sea-surface reflection and Earth curvature, into the joint beamforming design.} The differences between maritime and terrestrial communications are mainly attributed to the uniqueness of communication channels. Hereby, based on the special communication environment at sea, we model the shore-to-ship communication channel with two-ray path-loss model. Moreover, the TBS and the LEO satellite transmit dual-function signals, providing communication services for maritime users while also being able to jointly sensing the target. {To overcome the exponential complexity of traditional sensing, we propose an improved DE algorithm with an adaptive mutation and cross operator to efficiently locate the target.} Considering the huge resource consumption of traditional grid search and two-step positioning method secondary, we propose an algorithm to directly locate the target based on different evolution (DE) method . The proposed sensing algorithm jointly locate the position of target by utilizing the delay, elevation angle and azimuth angle of signal collected from the TBS and the LEO satellite. Then, we derived CRB as the measurement value of location accuracy, which is the determined bound for the variance of the unbiased estimator. Further, by jointly designing the beamformings of the TBS and the LEO satellite, we aim to maximize the sum rate of maritime users while meeting the localization accuracy requirements and transmit power constraints. The main contributions of this paper can be summarized as follows.

\begin{enumerate}
	
	\item We introduce a joint communication and sensing design for the ISTMS. The TBS and the LEO satellite provide wide-area network coverage for both near-shore users (NSUs) and off-shore users (OSUs), while simultaneously performing joint target sensing and localization.
	
	\item We propose a sensing algorithm based on DE method, which can improve location accuracy and reduce resource consumption. Further, we derive the closed-form expression of CRB as location performance metric.
	
	\item We propose a beamforming design algorithm to maximize the sum rate of maritime users by jointly designing the beamformings of the TBS and the LEO satellite, while meeting the localization accuracy and transmit power constraints.
	
\end{enumerate}

The rest of this paper is organized as follows. Section \uppercase\expandafter{\romannumeral2} presents the signal and maritime channel model. In section \uppercase\expandafter{\romannumeral3}, we introduce the joint communication and sensing design and proposed a sensing algorithm based on DE method. We also derive the performance metrics of communication and sensing. In section \uppercase\expandafter{\romannumeral4}, we jointly design the beamforming of the TBS and the LEO satellite to maximize the sum rate of maritime users, while meeting the location accuracy and transmit power constraints. Section \uppercase\expandafter{\romannumeral5} provides simulation results to verify effectiveness of proposed algorithms compared with the baseline algorithms. Finally, section \uppercase\expandafter{\romannumeral6} summarizes this paper.

\emph{Notations}: In this article, $\mathcal{N}(\mu,\sigma^2)$ and $\mathcal{CN}(\mu,\sigma^2)$ denote the Gaussian and complex Gaussian distribution with mean $\mu$ and variance $\sigma^2$, respectively. $J_1(\cdot)$ and $J_3(\cdot)$ denote the fist and third order of the first-kind of Bessel function, respectively. ${{\mathbb{C}}^{m\times n}}$ denotes the set of $m\times n$ dimensional complex matrices. Vectors and matrices are represented by bold letters. $\|\cdot\|_F$, $|\cdot|$, $(\cdot)^H$, $\|\cdot\|$ denote Frobenius norm of a matrix, absolute value, conjugate transpose and $L_2$-norm of a vector, respectively. $\text{tr}(\cdot)$ and $\text{Rank}(\cdot)$ represent the trace and rank of a matrix. {Further, the key symbols and their corresponding units used throughout this paper are summarized in Table \uppercase\expandafter{\romannumeral1}.}

\begin{table}[h]
	\small
	\centering
	\caption{List of Key Symbols}\label{Symbols}
	\begin{tabular}{|c|c|c|}
		\hline
		Symbol & Meaning  & Unit     \\\hline \hline
		$\mathbf{h}$   & channel gain  & dB \\\hline
		   $g_m$      &   large fading & dB \\\hline
		$\Gamma_{m}$   & satellite antenna gain & dBi \\\hline
		$\mathbf{x}$ & transmit signal & dB\\\hline
	    $\mathbf{w}$ and $\mathbf{v}$ & beamforming vector & dB\\\hline
		$P$      &  transmit power  & dBm\\\hline
		$y$  & received signal & dB\\\hline
		$R$   & data rate & $\text{bps/Hz}$\\\hline
		$\gamma$ & SINR & dB\\\hline
		$\widetilde{\mathbf{y}}$  & echo signal & dB\\\hline
		$\xi$  &  sensing channel gain & dB\\\hline
		$\mathbf{A}$ &  antenna array response& dB\\\hline
		$\mathbf{C}$  &  CRB& $\text{m}^2$\\ \hline
		$\mathbf{J}$ & Jacobian matrix& $\text{s}/\text{m}$ \\ \hline
		$\mathbf{F}$   & FIM  &$\text{m}^{-2}$\\ \hline
		$\boldsymbol{\rho}$ &  location vector& $\text{m}$\\\hline
	\end{tabular}
\end{table}

\section{System Model}
\begin{figure}[h]
	\centering
	\includegraphics[width=0.48\textwidth]{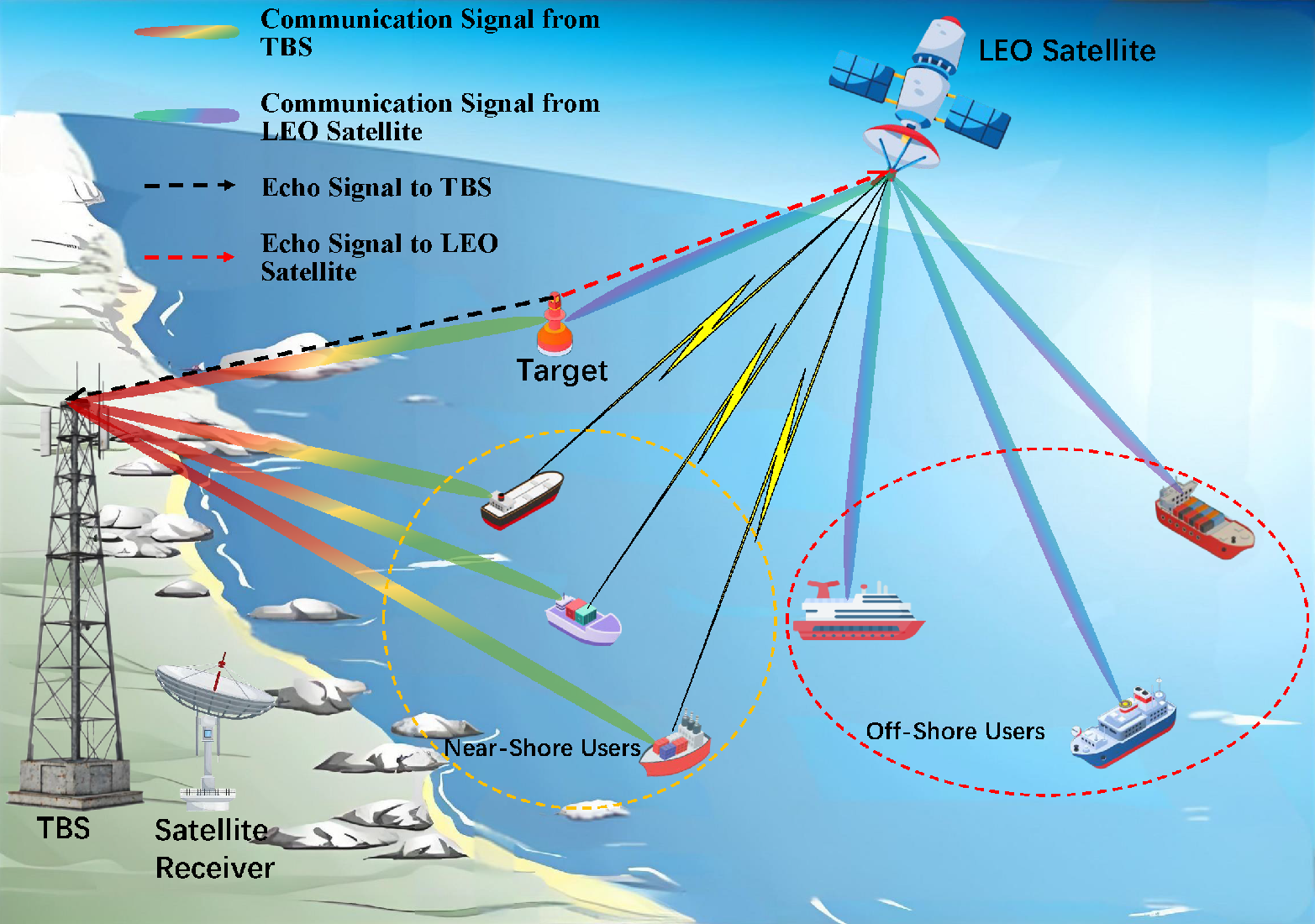} 
	\caption{System model of an integrated satellite-terrestrial maritime system.}
	\label{system}
\end{figure}

As shown in Fig. $\ref{system}$, we consider an integrated satellite-terrestrial maritime system (ISTMS), where a shore-based terrestrial base station (TBS) and a LEO satellite cooperatively provide communication services to various maritime users over the same frequency band while simultaneously performing target sensing \footnote{This paper assumes a single dominant reflection coefficient for the channel model, perfect CSI at both the TBS and the LEO satellite, and noise-free synchronization after Doppler compensation.}. Considering the sea surface obstruction caused by the curvature of the Earth, maritime users are usually divided into near-shore users (NSUs) and off-shore users (OSUs) based on the maximum effective distance of the TBS, which can be expressed as
\begin{equation}\label{dlos}
	\begin{aligned}
		d_{\text{LOS}}=\sqrt{h_r^{2}+2h_rR_e}+\sqrt{h_t^{2}+2h_tR_e},
	\end{aligned}
\end{equation}
where $R_e$ denotes the radius of the earth, $h_r$ and $h_t$ are the effective antenna heights of users and the TBS, respectively. Specifically, we define users whose distance from the TBS is less than $d_{\text{LOS}}$ as the NSUs, otherwise as OSUs. Accordingly, the TBS communicates with $K_1$ NSUs with a uniform planar array (UPA) of $M_1$ antennas, and the LEO satellite serves $K_2$ OSUs with a UPA of $M_2$ antennas. Due to size limitation, both NSUs and OSUs are equipped with a single antenna each. Meanwhile, the TBS and the LEO satellite sense the location of the target with the communication signals.

In the near-shore region, the shore-to-ship channel is mainly composed of a line of sight (LoS) component and a reflection component of the sea surface, following a two-ray model. {Compared to terrestrial environments, maritime channels typically feature fewer scatters and more dominant reflected paths. Thus, the two-ray model is adopted as a widely verified approach to accurately characterize these unique propagation characteristics \cite{TR}-\cite{28}}. Since the effects of scattering are considerably less compared to the primary components in the two-ray model, the impact caused by the diffuse reflection can be disregarded from the rough sea surface. Consequently, the channel between the TBS and the $i$-th NSU can be expressed as
\begin{equation}\label{h1}
	\begin{aligned}
		\mathbf{h}_{1,i}=\frac{\lambda }{2\pi d_i}\sin\bigg(\frac{2\pi h_th_r}{\lambda d_i}\bigg),
	\end{aligned}
\end{equation}
where $\lambda$ is the wavelength and $d_i$ denotes the distance between the $i$-th NSU and the TBS. Meanwhile, considering the signal propagation characteristics of the LEO satellite, the channel gain from the LEO satellite to the $m$-th OSU can be modeled as \cite{ym1}
\begin{equation}\label{h2}
	\begin{aligned}
		\mathbf{h}_{2,m}=g_{m}\bigg(\sqrt{\frac{1}{1+K_m}}\mathbf{h}_{2,m}^{\text{NLoS}}+\sqrt{\frac{K_m}{1+K_m}}\mathbf{h}_{2,m}^{\text{LoS}}\bigg),
	\end{aligned}
\end{equation}
where $K_m$ denotes the Rician factor, $\mathbf{h}_{2,m}^{\text{NLoS}}$ is the non-line of sight (NLoS) component of $\mathbf{h}_{2,m}$, which follows distribution $\mathcal{C}\mathcal{N}(0,1)$, and $\mathbf{h}_{2,m}^{\text{LoS}}$ is the LoS component of $\mathbf{h}_{2,m}$. Furthermore, $g_m$ represents the large scale fading of the channel, which can be expressed as
\begin{equation}\label{gm}
	\begin{aligned}
		g_{m}=\sqrt{\frac{\lambda^{2}}{(4\pi )^{2}D_m} \cdot \frac{G_m\Gamma_m}{r_m}  },
	\end{aligned}
\end{equation}
where $\frac{\lambda^{2}}{(4\pi )^{2}D_m}$ is the free space path loss with $D_m$ being the distance between the $m$-th OSU and the LEO satellite. $G_m$ denotes the $m$-th OSU's receive antenna gain and $r_m$ is the rain attenuation coefficient related to the $m$-th OSU whose power gain in dB $r_{m}^{\text{dB}}=20\log_{10}r_m$ follows log-normal random distribution $\ln(r_m^{\text{dB}})\sim \mathcal{N}(\mu_r, \sigma_r^2)$, with $u_r$ and $\sigma_r^2$ being the log-normal location and scale parameter, respectively. Moreover, $\Gamma_m$ is the $M_2$-dimensional satellite antenna gain, which can be modeled as \cite{cjh}
\begin{equation}\label{Pm}
	\begin{aligned}
		\Gamma_{m}=G_{\max}\bigg(\frac{J_1(\psi_m)}{2\psi_m}+36\frac{J_3(\psi_m)}{\psi_{m}^{3}}\bigg)^{2},
	\end{aligned}
\end{equation}
where $\psi_m=2.07\sin{\varphi_{\text{sm}}}/\sin{\varphi_{\text{3dB}}}$ with $\varphi_{\text{3dB}}$ and $\varphi_{\text{sm}}$ being the 3dB angle and the angle between the LEO satellite and the $m$-th OSU, respectively. $G_{\max}$ is the LEO satellite's maximum antenna gain.

In order to enhance the utilization efficiency of limited spectrum, the LEO satellite and the TBS construct and transmit the superimposed signals over the same very high frequency (VHF) band. In the communication mode, maritime users receive and decode information-bearing signals from the TBS and the LEO satellite. In the sensing mode, based on the bistatic radar principles, the TBS and the LEO satellite jointly process echo signals reflected from maritime targets to estimate their positions. Specifically, the transmit signal $\mathbf{x}_1$ from the TBS is modeled as \cite{ym3}
\begin{equation}\label{x1}
	\begin{aligned}
		\mathbf{x}_1(t)=\sum_{i=1}^{K_1}\mathbf{w}_is_{1,i}(t),
	\end{aligned}
\end{equation}
where $\mathbf{w}_i \in \mathbb{C}^{M_1 \times 1}$ is the transmit beamforming vector and $s_{1,i}$ is the information symbol with unit norm for the $i$-th NSU. Accordingly, the transmit signal $\mathbf{x}_2$ at the LEO satellite can be expressed as
\begin{equation}\label{x2}
	\begin{aligned}
		\mathbf{x}_2(t)=\sum_{m=1}^{K_2}\mathbf{v}_ms_{2,m}(t),
	\end{aligned}
\end{equation}
where $\mathbf{v}_m\in \mathbb{C}^{M_2 \times 1}$ is the transmit beamforming vector and $s_{2,m}$ denotes the normalized information symbol the $m$-th OSU. Besides, it is assumed that the all information symbols for a certain user are independent of each other and the information symbols for different users are uncorrelated. Therefore, the transmit power of the TBS and the LEO satellite can be computed as
\begin{equation}\label{p1}
	\begin{aligned}
		P_T=\sum_{i=1}^{K_1}|| \mathbf{w}_i ||^2,
	\end{aligned}
\end{equation}
and
\begin{equation}\label{p2}
	\begin{aligned}
		P_S=\sum_{m=1}^{K_2}|| \mathbf{v}_m ||^2.
	\end{aligned}
\end{equation}

In this section, we give the channel and signal models for the considered ISTMS. In the following, we design the joint communication and sensing scheme based on the same radio frequency signals.

\section{Joint Communication and Sensing Design}

In the proposed ISTMS, the TBS and the LEO satellite provide communication services to maritime users, while simultaneously performing target sensing. In the following, we propose a joint communication and sensing scheme according to the characteristics of the considered ISTMS.

\subsection{Communication Model}
Due to the wide beams of the LEO satellite, the NSUs communicating with the TBS are interfered by the LEO satellite. Consequently, the received signal at the $i$-th NSU can be written as
\begin{equation}\label{yi}
	\begin{aligned}
		y_{1,i}(t)&=\mathbf{h}_{1,i}\mathbf{x}_1(t-\tau_{1,i})e^{j2\pi v_{1,i}t}+\\
		&\quad\quad\quad\quad\hat{\mathbf{h}}_{2,i}\mathbf{x}_2(t-\tau_{1,i})e^{j2\pi v_{1,i}t}+n_{1,i}(t)  \\
		&=\underbrace{\mathbf{h}_{1,i}\mathbf{w}_i s_{1,i}(t-\tau_{1,i})e^{j2\pi v_{1,i}t}}_{\text{Desired  Signal}}+\\
		&\quad \underbrace{\sum_{j=1,j\neq i}^{K_1} \mathbf{h}_{1,i}\mathbf{w}_js_{1,j}(t-\tau_{1,i})e^{j2\pi v_{1,i}t}}_{\text{Adjacent Channel Interference}}+ \\
		&\quad \underbrace{\sum_{m=1}^{K_2} \hat{\mathbf{h}}_{2,i}\mathbf{v}_ms_{2,m}(t-\tau_{1,i})e^{j2\pi v_{1,i}t}}_{\text{LEO Satellite Interference}}+\underbrace{n_{1,i}(t)}_{\text{AWGN}},
	\end{aligned}
\end{equation}
where $\hat{\mathbf{h}}_{2,i}$ represents the channel from the LEO satellite to the $i$-th NSU, and $n_{1,i}$ is the additive white Gaussian noise (AWGN) at the $i$-th NSU with variance $\sigma_{1,i}^{2}$. Besides, $v_{1,i}$ and $\tau_{1,i}$ are the Doppler shift and propagation delay associated with the channel from the TBS to the $i$-th NSU, which can be estimated and compensated by the time-frequency synchronization in advance at the receiver \cite{ym2}-\cite{DP2}. Based on the maximum effective distance of the TBS, the OSUs can only receive signals from the LEO satellite. Thus, the received signal at the $m$-th OSU can be formulated as
\begin{equation}\label{ym}
	\begin{aligned}
		y_{2,m}(t)&=\mathbf{h}_{2,m}\mathbf{x}_2(t-\tau_{2,m})e^{j2\pi v_{2,m}t}+n_{2,m}(t) \\
		&=\underbrace{\mathbf{h}_{2,m}\mathbf{v}_m s_{2,m}(t-\tau_{2,m})e^{j2\pi v_{2,m}t}}_{\text{Desired Signal}}+\\
		&\quad \underbrace{ \sum_{k=1,k\neq m}^{K_2}\mathbf{h}_{2,m}\mathbf{v}_ks_{2,k}(t-\tau_{2,m})e^{j2\pi v_{2,m}t}}_{\text{Adjacent Channel Interference}}+\underbrace{n_{2,m}(t)}_{\text{AWGN}},
	\end{aligned}
\end{equation}
where $n_{2,m}$ is the AWGN at the $m$-th OSU with variance $\sigma_{2,m}^{2}$, and $v_{2,m}$ and $\tau_{2,m}$ denote the Doppler shift and delay associated with the channel from the LEO satellite to the $m$-th OSU, which can also be estimated and compensated at the receiver. {Specifically, the TBS and LEO satellite achieve synchronization via GNSS pulse-per-second signals, while pre-processed sensing data are aggregated at the TBS through high-speed feeder links for centralized coordinate transformation and spatial alignment.} For ease of presentation, we omit the index in the following.

In order to evaluate the communication quality of maritime users, we choose the data transmission rate as the communication performance metric. Specifically, the data transmission rate of the $i$-th NSU can be expressed as
\begin{equation}\label{R1}
	\begin{aligned}
		R_{1,i} = \log_2(1+\gamma_{1,i}),
	\end{aligned}
\end{equation}
where $\gamma_{1,i}$ denotes the signal to interference plus noise ratio (SINR) at the $i$-th NSU, which is given by
\begin{equation}\label{SINR1}
	\begin{aligned}
		\gamma_{1,i}=\frac{|\mathbf{h}_{1,i}\mathbf{w}_i|^2}{\sum_{j=1,j\ne   i}^{K_1}|\mathbf{h}_{1,i}\mathbf{w}_j|^2+\sum_{m=1}^{K_2}|\hat{\mathbf{h}}_{2,i}\mathbf{v}_m|^2+\sigma_{1,i}^{2}}.
	\end{aligned}
\end{equation}
Similarly, the data transmission rate of the $m$-th OSU can be expressed as
\begin{equation}\label{R2}
	\begin{aligned}
		R_{2,m} = \log_2(1+\gamma_{2,m}),
	\end{aligned}
\end{equation}
where $\gamma_{2,m}$ represents the SINR at the $m$-th OSU, which can be given by
\begin{equation}\label{SINR2}
	\begin{aligned}
		\gamma_{2,m}=\frac{|\mathbf{h}_{2,m}\mathbf{v}_m|^2}{\sum_{k=1,k\ne   m}^{K_2}|\mathbf{h}_{2,m}\mathbf{v}_k|^2+\sigma_{2,m}^{2}}.
	\end{aligned}
\end{equation}

\subsection{Sensing Model}
To monitor the maritime target and improve navigation safety, we aim to locate the target and broadcast its location information to the maritime users. Let $\mathbf{q}_1=(q_1^x,q_1^y,z_1^z)$, $\mathbf{q}_2=(q_2^x,q_2^y,q_2^z)$ and $\mathbf{p}=(p_x,p_y,p_z)$ represent the location vectors of the TBS, the LEO satellite and the target in Cartesian coordinate system that the origin set as the center of the Earth, respectively. In the proposed ISTMS, the transmit signals of the TBS and the LEO satellite are reflected by the target. Hence, the received echo signal at the TBS can be expressed as{ \footnote{While real-world factors like sea clutter and RCS fluctuations exist, the two-ray model is adopted as it effectively captures the dominant propagation mechanism in maritime environments. Specifically, at VHF frequencies, the large wavelength relative to sea surface roughness makes specular reflection dominant.}}
\begin{equation}\label{ytbs}
	\begin{aligned} \widetilde{\mathbf{y}}_T=\alpha\xi_1\mathbf{A}_1(\theta_1,\varphi_1)\widetilde{\mathbf{x}}_1+\alpha\xi_2\mathbf{A}_2(\theta_2,\varphi_2)\widetilde{\mathbf{x}}_2+\widetilde{\mathbf{c}}_1,
	\end{aligned}
\end{equation}
where $\alpha$ is the reflection coefficient of the target. $\xi_1=|\mathbf{h}_1\mathbf{h}_1^H|$ and $\xi_2=|\mathbf{h}_1\mathbf{h}_2^H|$ represent the sensing channel gain from the TBS and the LEO satellite through target reflection to the TBS, respectively, and $\widetilde{\mathbf{c}}_1$ denotes the noise vector at the TBS. Further, $\widetilde{\mathbf{x}}_1 = [\mathbf{x}_1e^{j2\pi v_1t},\mathbf{0}]\mathbf{J}(\tau_1)$ and $\widetilde{\mathbf{x}}_2 = [\mathbf{x}_2e^{j2\pi v_2t},\mathbf{0}]\mathbf{J}(\tau_2)$ represent the delayed and Doppler-shift counterpart of $\mathbf{x}_1$ and $\mathbf{x}_2$,  where $v_1$ and $v_2$ are the Doppler shift frequency from the TBS and the LEO satellite. Meanwhile, $\mathbf{J}(\tau_1)$ and $\mathbf{J}(\tau_2)$ represent the time-domain shifting matrix with $\tau_1$ and $\tau_2$ being the delay from the TBS and the LEO satellite, respectively, which can be expressed as \cite{cj2}
\begin{equation}
	\begin{aligned}
		\mathbf{J}(\tau_k)=
		\begin{bmatrix}
			\underbrace{0 \dots 0}_{\tau_k} &1& \cdots& 0\\
			0 \dots  0  & 0 & \ddots &  \vdots \\
			\vdots & 0 & \cdots & 1\\
			\vdots &\vdots &\vdots&\vdots\\
			0 \dots 0 &0 & \cdots& 0
		\end{bmatrix}, k = 1, 2.
	\end{aligned}
\end{equation}

Moreover, $\mathbf{A}_1(\theta_1,\varphi_1)=\mathbf{a}_1^r(\theta_1,\varphi_1)\mathbf{a}_1^t(\theta_1,\varphi_1)$ and $\mathbf{A}_2(\theta_2,\varphi_2)=\mathbf{a}_2^r(\theta_2,\varphi_2)\mathbf{a}_2^t(\theta_2,\varphi_2)$ denote the antenna array response of the link from the TBS to the target and from the LEO satellite to the target, respectively. $\theta_1$ and $\varphi_1$ represent the elevation and azimuth angle at the TBS, and $\theta_2$ and $\varphi_2$ denote the elevation and azimuth angle at the LEO satellite. Specifically, $\mathbf{a}^r$ and $\mathbf{a}^t$ have the same structure and can be expressed as
\begin{equation}
	\begin{aligned}
		&\mathbf{a}_k^r(\theta_k,\varphi_k)=\mathbf{a}_k^t(\theta_k,\varphi_k)=\\
		&\quad\bigg(\frac{1}{\sqrt{M_k^x}}[1,e^{j\frac{2\pi d}{\lambda}\sin\theta_k\cos\varphi_k},\dots,e^{j\frac{2\pi d}{\lambda}(M_k^x-1)\sin\theta_k\cos\varphi_k}]\bigg)\otimes\\
		&\quad\quad\bigg(\frac{1}{\sqrt{M_k^z}}[1,e^{j\frac{2\pi d}{\lambda}\cos\theta_k},\dots,e^{j\frac{2\pi d}{\lambda}(M_k^z-1)\cos\theta_k}]\bigg)\\
		&\quad\quad\quad=\frac{1}{M_k}[1,e^{j\frac{2\pi d}{\lambda}(\sin\theta_k\cos\varphi_k+\cos \theta_k)},\dots,\\
		&\quad\quad\quad\quad e^{j\frac{2\pi d}{\lambda}((M_k^x-1)\sin\theta_k\cos\varphi_k+(M_k^z-1)\cos \theta_k)}],
	\end{aligned}
\end{equation}
where $k={1,2}$, $M_k=M_k^x\times M_k^z$ with $M_k^x$ and $M_k^z$ being the number of antennas on the x-axis and z-axis of the UPA at the TBS and the LEO satellite, respectively, and $d$ is the adjacent antenna spacing. Similarly, the echo signal at the LEO satellite can be modeled as
\begin{equation}\label{ysat}
	\begin{aligned} \widetilde{\mathbf{y}}_S=\alpha\xi_3\mathbf{A}_2(\theta_2,\varphi_2)\widetilde{\mathbf{x}}_2+\alpha\xi_2\mathbf{A}_1(\theta_1,\varphi_1)\widetilde{\mathbf{x}}_1+\widetilde{\mathbf{c}}_2,
	\end{aligned}
\end{equation}
where $\xi_3=|\mathbf{h}_2\mathbf{h}_2^H|$ is the sensing channel gain from the LEO satellite through target reflection to the LEO satellite, and $\widetilde{\mathbf{c}}_2$ is the noise vector at the LEO satellite.

To estimate the angel of arrive (AoA) of the echo signal, the maximum likelihood estimation (MLE) algorithm can be used. For a given delay, we can get the likelihood function of $\widetilde{\mathbf{y}}_T$ with new auxiliary angle vectors $\boldsymbol{\eta}_1 = [\theta_1,\varphi_1]$ and $\boldsymbol{\eta}_2 = [\theta_2,\varphi_2]$ as
\begin{equation}\label{L1}
	\begin{aligned}
		&L(\widetilde{\mathbf{y}}_T;\boldsymbol{\eta}_1,\alpha)=\frac{1}{\sqrt{(2\pi)^{M_1}\sigma_{1}^2}} \\
		& \exp \bigg[ -\frac{1}{\sigma_{1}^2} ||\widetilde{\mathbf{y}}_T- \alpha\xi_1\mathbf{A}_1(\boldsymbol{\eta}_1)\widetilde{\mathbf{x}}_1-\alpha\xi_2\mathbf{A}_2(\boldsymbol{\eta}_2)\widetilde{\mathbf{x}}_2||^2    \bigg].
	\end{aligned}
\end{equation}
Due to the reflection coefficient $\alpha$ and angle vector $\boldsymbol{\eta}_1$ being both unknown parameters, the MLE of $(\boldsymbol{\eta}_1, \alpha)$ is written as
\begin{equation}\label{b}
	\begin{aligned}
		&(\boldsymbol{\eta}_1^*,\alpha^*)=\mathop{\arg\max}\limits_{\boldsymbol{\eta}_1,\alpha} L(\widetilde{\mathbf{y}}_T;\boldsymbol{\eta}_1,\alpha) \\
		& \quad = \mathop{\arg\min}\limits_{\boldsymbol{\eta}_1,\alpha} ||\widetilde{\mathbf{y}}_T- \alpha\xi_1\mathbf{A}_1(\boldsymbol{\eta}_1)\widetilde{\mathbf{x}}_1-\alpha\xi_2\mathbf{A}_2(\boldsymbol{\eta}_2)\widetilde{\mathbf{x}}_2||^2.
	\end{aligned}
\end{equation}
For given $\boldsymbol{\eta}_1$, the MLE of $\alpha$ can be determined as
\begin{equation}\label{e}
	\begin{aligned}
		&\alpha^*= \mathop{\arg\min}\limits_{\alpha} ||\widetilde{\mathbf{y}}_T- \alpha\xi_1\mathbf{A}_1(\boldsymbol{\eta}_1)\widetilde{\mathbf{x}}_1-\alpha\xi_2\mathbf{A}_2(\boldsymbol{\eta}_2)\widetilde{\mathbf{x}}_2||^2\\
		&\quad=\frac{\widetilde{\mathbf{y}}_T^H\big(\xi_1\mathbf{A}_1(\boldsymbol{\eta}_1)\widetilde{\mathbf{x}}_1+\xi_2\mathbf{A}_2(\boldsymbol{\eta}_2)\widetilde{\mathbf{x}}_2\big)}{||\xi_1\mathbf{A}_1(\boldsymbol{\eta}_1)\widetilde{\mathbf{x}}_1+\xi_2\mathbf{A}_2(\boldsymbol{\eta}_2)\widetilde{\mathbf{x}}_2||^2}.
	\end{aligned}
\end{equation}
Then, substituting (\ref{e}) into (\ref{b}), the MLE of $\boldsymbol{\eta}_1$ can be computed as
\begin{equation}\label{a}
	\begin{aligned}
		&\boldsymbol{\eta}_1^*=\mathop{\arg\min}\limits_{\boldsymbol{\eta}_1} ||\widetilde{\mathbf{y}}_T- \alpha^*(\xi_1\mathbf{A}_1(\boldsymbol{\eta}_1)\widetilde{\mathbf{x}}_1+\xi_2\mathbf{A}_2(\boldsymbol{\eta}_2)\widetilde{\mathbf{x}}_2)||^2\\
		&=\mathop{\arg\min}\limits_{\boldsymbol{\eta}_1} \bigg(   ||\widetilde{\mathbf{y}}_T^2||-  \frac{|\widetilde{\mathbf{y}}_T^H\big(\xi_1\mathbf{A}_1(\boldsymbol{\eta}_1)\widetilde{\mathbf{x}}_1+\xi_2\mathbf{A}_2(\boldsymbol{\eta}_2)\widetilde{\mathbf{x}}_2\big)|^2}{||\xi_1\mathbf{A}_1(\boldsymbol{\eta}_1)\widetilde{\mathbf{x}}_1+\xi_2\mathbf{A}_2(\boldsymbol{\eta}_2)\widetilde{\mathbf{x}}_2||^2}   \bigg).
	\end{aligned}
\end{equation}

When dealing with the problem of maximum likelihood function, the grid search method is a common approach. However, from (\ref{a}), we can find that when the number of antennas and the dimension of the matrix increase, the computational complexity of the grid search method increases exponentially. In order to obtain a more accurate global optimal solution, it usually chooses a small step distance for exhaustive solution, which undoubtedly leads to an exponential increase in the solution time and consumed resources. Therefore, the grid search approach cannot meet the real-time requirements of maritime systems.

To overcome this challenge, we adopt the differential evolution (DE) algorithm with low-complexity and fast-convergence characteristics. Specifically, the DE algorithm is a population-based evolutionary algorithm that is originated from the process of cooperation and competition among individuals within a population. In the DE algorithm, each individual represents a candidate solution to the problem. Before each iteration, it performs a mutation operation first to obtain the new individuals. Further, through cross operation, the new individual is crossed with the corresponding individual of the parent generation. Then, by performing the selection, the better individual is selected and retained to the next generation. After several iterations, the optimal individual of the population is the solution of the problem.

To be specific, we let the $N_p$ represent the number of population, $G_{\max}$ is the maximum number of iterations, $g$ denotes the iteration index. $p_i^{(g)}$, $v_i^{(g)}$ and $u_i^{(g)}$ is the initial population individuals, the new individuals obtained after mutation and the new individuals selected through cross-selection, respectively. In the DE algorithm, population individuals perform intelligent search through information exchange and cooperation. Furthermore, the MLE of $\boldsymbol{\eta}_1$ is used as the fitness function to characterize the accuracy of angle estimation. By continuously updating population individuals, the angle sensing algorithm gradually approaches the global optimum $\boldsymbol{\eta}_1^*$. In order to obtain a more accurate global optimal solution, we make some changes to the mutation and cross operator compared to the traditional DE algorithm, respectively. Specifically, we set an adaptive and dynamically changing mutation operator, which can be expressed as
\begin{equation}\label{Fnew}
	\begin{aligned}
		F = F_0 \times 2^{\lambda}
	\end{aligned}
\end{equation}
where $F_0$ denotes the initial mutation operator and  $\lambda=\exp(\frac{1-g}{G_{\max}+1-g})$. In addition, we also set a random cross operator, which can be modeled as
\begin{equation}\label{Cnew}
	\begin{aligned}
		C_R=0.5\times (1+r_1)
	\end{aligned}
\end{equation}
where $r_1$ represents a random number ranging from 0 to 1, respectively. With (\ref{Fnew}) and (\ref{Cnew}), the proposed algorithm has greater randomness, which is helpful to escape from the local optimal solution. {Specifically, Algorithm 1 conducts a joint search over the unified parameter space $\boldsymbol{\eta} = [\boldsymbol{\eta}_1,\boldsymbol{\eta}_2]$ where the candidate vectors for both TBS and LEO angle parameters evolve simultaneously to minimize the global cost function and resolve potential coupling dependencies.} Last, we use a simple matched filtering algorithm to estimate the delay parameter. The detailed procedure of the algorithm are listed in Algorithm 1.

\begin{algorithm}[!h]
	\caption{DE-based Direct Location Sensing Algorithm}
	\label{alg:AOA}
	\renewcommand{\algorithmicrequire}{\textbf{Input:}}
	\renewcommand{\algorithmicensure}{\textbf{Output:}}
	\begin{algorithmic}[1]
		\REQUIRE $\widetilde{\mathbf{y}}_T$, $\mathbf{w}_i$, $\mathbf{v}_m$, $\mathbf{h}_{1,i}$, $\mathbf{H}_{2,m}$, $N_p$, $F_0$, $C_R$
		\STATE $\mathbf{Initialization:}$ Set initial population $\{ p_i^{(0)}  \}^{N_p}_{i=1}$, iteration index $g=0$, maximum iterations $G_{\max}$, random numbers $r_1 \in [0,1]$ and $r_2 \in [0,1]$, dimension $d_{\mathrm{rand}}$, initial cross operator $F_0$,
		\FOR{all $\tau_1$ in searching space do }
		\WHILE{$\mathbf{while}$ $g<G_{\max}$}
		\FOR{$i=1:N_p$}
		\STATE generate mutation vector :$v_i^{(g)} = p_{r_1}^{(g)} + F\times (p_{r_2}^{(g)}-p_{r_3}^{(g)})$, $r_1,r_2,r_3 \neq i$
		\IF{$r_2<C_R$ or $d<d_{\mathrm{rand}}$}
		\STATE $u_{i,d}^{(g)}=v_{i,d}^{(g)}$
		\ELSE
		\STATE  $u_{i,d}^{(g)}=p_{i,d}^{(g)}$
		\ENDIF
		\STATE Calculate fitness function : \\
		$F_1(p_i^{(g)})=||\widetilde{\mathbf{y}}_T^2||-  \frac{|\widetilde{\mathbf{y}}_T^H\big(\xi_1\mathbf{A}_1(p_i^{(g)})\widetilde{\mathbf{x}}_1+\xi_2\mathbf{A}_2(\eta_2)\widetilde{\mathbf{x}}_2\big)|^2}{||\xi_1\mathbf{A}_1(p_i^{(g)})\widetilde{\mathbf{x}}_1+\xi_2\mathbf{A}_2(\eta_2)\widetilde{\mathbf{x}}_2||^2}$
		\IF{$F_1(u_i^{(g)})>F_1(p_i^{(g)})$}
		\STATE Update individual $p_i^{(g+1)}=u_i^{(g)}$
		\ELSE
		\STATE $p_i^{(g+1)}=p_i^{(g)}$
		\ENDIF
		\ENDFOR
		\STATE Update iteration index : $g=g+1$
		\ENDWHILE
		\STATE Obtain estimated angle $\boldsymbol{\eta}_1^*$
		\STATE Compute $\mathop{\arg\max}\limits_{\tau_1}|\widetilde{\mathbf{y}}_T(\boldsymbol{\eta}_1^*)^H\widetilde{\mathbf{y}}_T(\boldsymbol{\eta}_1)|$ to obtain $\tau_1^*$
		\ENDFOR
		\STATE Obtain estimated angle $\boldsymbol{\eta}_1^*$, $\tau_1^*$
		\ENSURE $\boldsymbol{\eta}_1^* = (\theta_1^*,\varphi_1^*)$ and $\tau_1^*$    
	\end{algorithmic}
\end{algorithm}

\begin{figure}[t]
	\centering
	\includegraphics[width=0.52\textwidth]{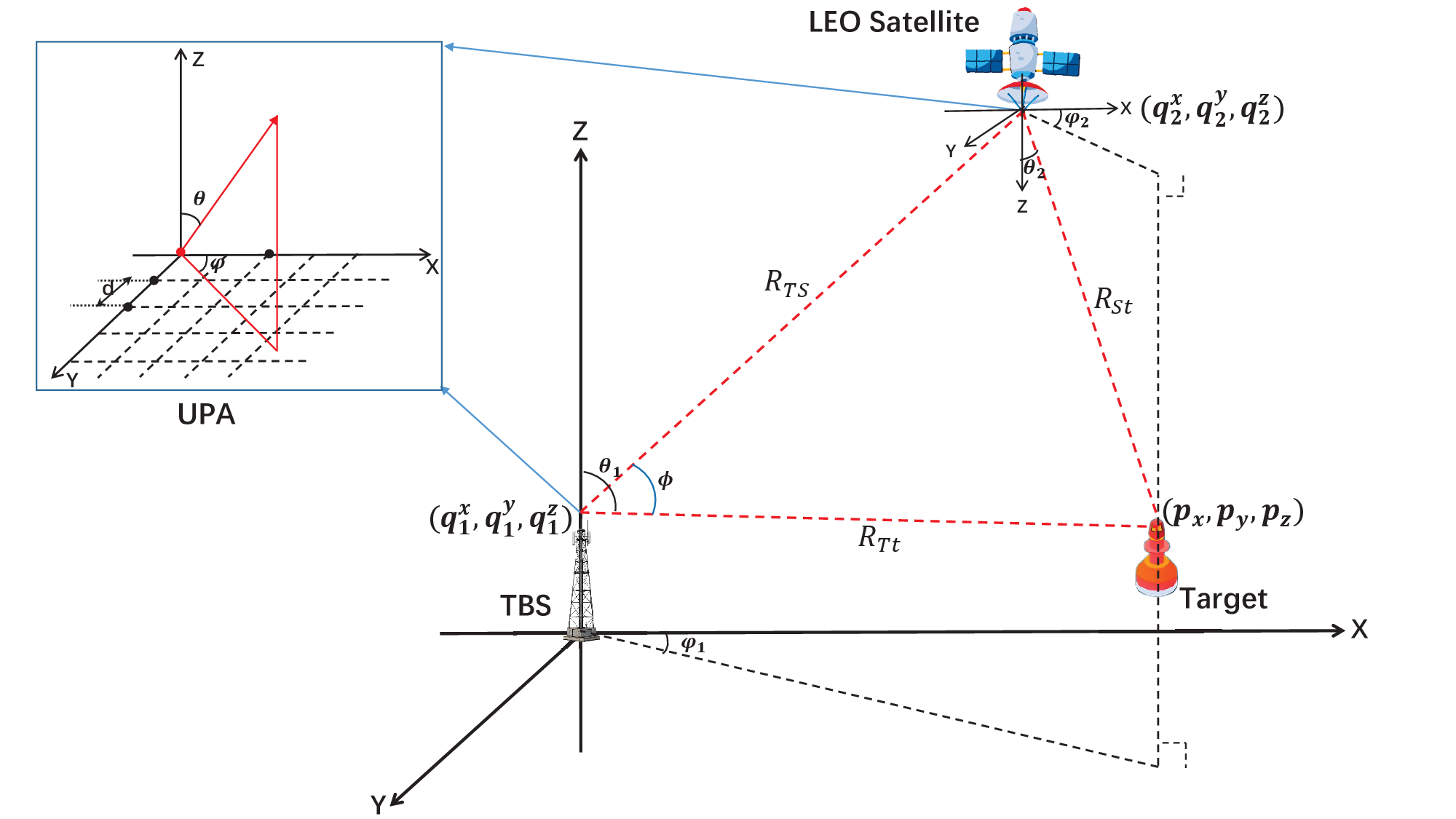} 
	\caption{Geometrical model of the integrated terrestrial-satellite maritime communication system.}
	\label{Gr}
\end{figure}

Similarly, the MLE of $\boldsymbol{\eta}_2$ can be expressed as
\begin{equation}
	\begin{aligned}
		&\boldsymbol{\eta}_2^*=\mathop{\arg\min}\limits_{\boldsymbol{\eta}_2} ||\widetilde{\mathbf{y}}_S- \alpha^*(\xi_3\mathbf{A}_2(\boldsymbol{\eta}_2)\widetilde{\mathbf{x}}_2+\xi_2\mathbf{A}_1(\boldsymbol{\eta}_1)\widetilde{\mathbf{x}}_1)||^2\\
		&=\mathop{\arg\min}\limits_{\boldsymbol{\eta}_2} \bigg(||\widetilde{\mathbf{y}}_S^2||-  \frac{|\widetilde{\mathbf{y}}_S^H\big(\xi_3\mathbf{A}_2(\boldsymbol{\eta}_2)\widetilde{\mathbf{x}}_2+\xi_2\mathbf{A}_1(\boldsymbol{\eta}_1)\widetilde{\mathbf{x}}_1\big)|^2}{||\xi_3\mathbf{A}_2(\boldsymbol{\eta}_2)\widetilde{\mathbf{x}}_2+\xi_2\mathbf{A}_1(\boldsymbol{\eta}_1)\widetilde{\mathbf{x}}_1||^2}   \bigg),
	\end{aligned}
\end{equation}
which can also be obtained by the proposed Algorithm 1. Then, the LEO satellite transmits the initially processed sensing parameters to the TBS, which aggregates heterogeneous sensing parameters from the LEO satellite and the TBS. Furthermore, as shown in Fig. $\ref{Gr}$, we can find the mapping relationships between angles and distances and the target position. Firstly, according to geometric locations, the elevation angle and azimuth angle of the target with respect to the TBS and the LEO satellite can be computed as
\begin{equation}\label{theta}
	\begin{aligned}
		\theta_k = \arccos \frac{p_z-q_k^z}{\sqrt{(q_k^x-p_x)^2+(q_k^y-p_y)^2+(q_k^z-p_z)^2}},
	\end{aligned}
\end{equation}
and
\begin{equation}\label{varphi}
	\begin{aligned}
		\varphi_k =
		\begin{cases}
			\arctan(\frac{p_y-q_k^y}{p_x-q_k^x})  \quad&\text{if} \quad p_x > q_k^x\\
			\arctan(\frac{p_y-q_k^y}{p_x-q_k^x})-\pi \quad \  &\text{if} \quad p_x < q_k^x,\  p_y < q_k^y\\
			\arctan(\frac{p_y-q_k^y}{p_x-q_k^x})+\pi \quad \  &\text{if} \quad p_x < q_k^x,\  p_y \geq q_k^y\\
			-\frac{\pi}{2}  &\text{if} \quad p_x = q_k^x,\  p_y < q_k^y\\
			+\frac{\pi}{2}  &\text{if} \quad p_x = q_k^x,\  p_y \geq q_k^y
		\end{cases}.
	\end{aligned}
\end{equation}
where $k = 1, 2$. Secondly, the distance from the TBS to the target can be calculated as
\begin{equation}\label{RTt}
	\begin{aligned}
		R_{Tt}=\frac{\nu \tau_1^*}{2}=\sqrt{(p_x-q_1^x)^2+(p_y-q_1^y)^2+(p_z-q_1^z)^2},
	\end{aligned}
\end{equation}
where $\nu$ denotes the propagation speed of electromagnetic wave. Lastly, the distance between the LEO satellite and the target can be computed as
\begin{equation}\label{RSt}
	\begin{aligned}
		R_{St}&=\frac{\nu \tau_2^*}{2}=\sqrt{(p_x-q_2^x)^2+(p_y-q_2^y)^2+(p_z-q_2^z)^2},
	\end{aligned}
\end{equation}
Then, the TBS collaboratively processes all the angle and delay information from the LEO satellite and the TBS. According to the mapping relationship equations (\ref{theta}), (\ref{varphi}), (\ref{RTt}) and (\ref{RSt}), the three-dimensional coordinate position of the target can be calculated at the TBS.

Further, we choose the CRB to analyze the performance of the proposed location sensing algorithm, which provides a lower bound for the mean squared error (MSE). Specifically, CRB represents the theoretical limit for parameter estimation, which is given by
\begin{equation}\label{C1}
	\begin{aligned}
		\mathbf{C}_1=(\mathbf{J}_1^T\mathbf{F}_{\boldsymbol{\rho}_1}\mathbf{J}_1)^{-1}
	\end{aligned}
\end{equation}
and
\begin{equation}\label{C2}
	\begin{aligned}
		\mathbf{C}_2=(\mathbf{J}_2^T\mathbf{F}_{\boldsymbol{\rho}_2}\mathbf{J}_2)^{-1},
	\end{aligned}
\end{equation}
where  $\boldsymbol{\rho}_1=[\theta_1,\varphi_1,\tau_1]$ and $\boldsymbol{\rho}_2=[\theta_2,\varphi_2,\tau_2]$ denote the sensing vector of the TBS and the LEO satellite, respectively. $\mathbf{F}_{\boldsymbol{\rho}_1}$ and $\mathbf{F}_{\boldsymbol{\rho}_2}$ represent the Fisher information matrix of location vector $\boldsymbol{\rho}_1$ and $\boldsymbol{\rho}_2$, respectively. Further, $\mathbf{J}_1$ and $\mathbf{J}_2$ are the Jacobian matrix of the linear mapping relation from $\boldsymbol{\rho}_1$ and $\boldsymbol{\rho}_2$ to $\mathbf{p}$, respectively, which can be computed by taking partial derivation. To be specific, $\mathbf{J}_k$ can be expressed as
\begin{equation}\label{J}
	\begin{aligned}
		\mathbf{J}_k=\frac{\partial \boldsymbol{\rho}_k}{\partial \mathbf{p}}=\bigg[\frac{\partial \boldsymbol{\rho}_k}{\partial p_x},\frac{\partial \boldsymbol{\rho}_k}{\partial p_y},\frac{\partial \boldsymbol{\rho}_k}{\partial p_z}\bigg],
	\end{aligned}
\end{equation}
where $\forall k=1,2$, $\frac{\partial \boldsymbol{\rho}_1}{\partial p_x}=[\frac{\partial \theta_1}{\partial p_x},\frac{\partial \varphi_1}{\partial p_x},\frac{\partial \tau}{\partial p_x}]$ and $\frac{\partial \boldsymbol{\rho}_2}{\partial p_x}=[\frac{\partial \theta_2}{\partial p_x},\frac{\partial \varphi_2}{\partial p_x}]$. Besides, $\frac{\partial \boldsymbol{\rho}_k}{\partial p_y}$ and $\frac{\partial \boldsymbol{\rho}_k}{\partial p_z}$ have the similar form. By taking the partial derivation, $\frac{\partial \theta_k}{\partial \mathbf{p}}$ can be calculated as
\begin{equation}
	\begin{aligned}\label{1}
		\frac{\partial \theta_k}{\partial p_x}=\frac{(p_z-q_k^z)(p_x-q_k^x)[(p_x-q_k^x)^2+(p_y-q_k^y)^2]^{-\frac{1}{2}}}{(p_x-q_k^x)^2+(p_y-q_k^y)^2+(p_z-q_k^z)^2},
	\end{aligned}
\end{equation}

\begin{equation}
	\begin{aligned}
		\frac{\partial \theta_k}{\partial p_y}=\frac{(p_z-q_k^z)(p_y-q_k^y)[(p_x-q_k^x)^2+(p_y-q_k^y)^2]^{-\frac{1}{2}}}{(p_x-q_k^x)^2+(p_y-q_k^y)^2+(p_z-q_k^z)^2},
	\end{aligned}
\end{equation}
and
\begin{equation}
	\begin{aligned}
		\frac{\partial \theta_k}{\partial p_z}=-\frac{[(p_x-q_k^x)^2+(p_y-q_k^y)^2]^{\frac{1}{2}}}{(p_x-q_k^x)^2+(p_y-q_k^y)^2+(p_z-q_k^z)^2}.
	\end{aligned}
\end{equation}
Similarly, $\frac{\partial \varphi_k}{\partial \mathbf{p}}$ can be calculated as
\begin{equation}
	\begin{aligned}
		\frac{\partial \varphi_k}{\partial p_x}=\frac{q_k^y-p_y}{(p_x-q_k^x)^2+(p_y-q_k^y)^2},
	\end{aligned}
\end{equation}

\begin{equation}
	\begin{aligned}
		\frac{\partial \varphi_k}{\partial p_y}=\frac{p_x-q_k^x}{(p_x-q_k^x)^2+(p_y-q_k^y)^2},
	\end{aligned}
\end{equation}
and
\begin{equation}
	\begin{aligned}
		\frac{\partial \varphi_k}{\partial p_z}=0.
	\end{aligned}
\end{equation}
According to (\ref{RTt}) and (\ref{RSt}), $\frac{\partial \tau_k}{\partial \mathbf{p}}$ can be calculated as
\begin{equation}
	\begin{aligned}
		\frac{\partial \tau_k}{\partial p_x}=\frac{2|p_x-q_k^x|}{\nu\sqrt{(p_x-q_k^x)^2+(p_y-q_k^y)^2+(p_z-q_k^z)^2}},
	\end{aligned}
\end{equation}

\begin{equation}
	\begin{aligned}
		\frac{\partial \tau_k}{\partial p_y}=\frac{2|p_y-q_k^y|}{\nu\sqrt{(p_x-q_k^x)^2+(p_y-q_k^y)^2+(p_z-q_k^z)^2}},
	\end{aligned}
\end{equation}
and
\begin{equation}
	\begin{aligned}\label{2}
		\frac{\partial \tau_k}{\partial p_z}=\frac{2|p_z-q_k^z|}{\nu\sqrt{(p_x-q_k^x)^2+(p_y-q_k^y)^2+(p_z-q_k^z)^2}}.
	\end{aligned}
\end{equation}

Lastly, by substituting equations (\ref{1})-(\ref{2}) into (\ref{J}), we can calculate the Jacobian matrix $\mathbf{J}_1$ and $\mathbf{J}_2$. In order to derive a closed-form expression of CRB from $\mathbf{F}_{\boldsymbol{\rho}_k}$, we denote $\boldsymbol{\xi}_1=[\boldsymbol{\rho}_1,\alpha]$ and $\boldsymbol{\xi}_2=[\boldsymbol{\rho}_2,\alpha]$ as the unknown parameters of the TBS and the LEO satellite to estimate, respectively. Then, the FIM for $\boldsymbol{\xi}_k$ can be expressed as
\begin{equation}\label{F}
	\begin{aligned}
		\mathbf{F}_{\boldsymbol{\xi}_k} =
		\begin{bmatrix}
			 \mathbf{F}_{\boldsymbol{\rho}_k \boldsymbol{\rho}_k}  & \mathbf{F}_{\boldsymbol{\rho}_k \alpha}\\
			 \mathbf{F}_{\boldsymbol{\rho}_k \alpha}^T     &  \mathbf{F}_{\alpha \alpha}
		\end{bmatrix},
	\end{aligned}
\end{equation}
where
\begin{equation}\label{Fij}
	\begin{aligned}
		[\mathbf{F}_{\boldsymbol{\xi}_k}]_{i,j} =\frac{2}{\sigma_k^2} \text{Re} \bigg\{ \frac{\partial \mathbf{u}_k^H}{\partial [\boldsymbol{\xi}_k]_i}  \frac{\partial \mathbf{u}_k}{\partial [\boldsymbol{\xi}_k]_j} \bigg\}, \forall k=1,2
	\end{aligned}
\end{equation}
with $\mathbf{u}_1=\alpha\xi_1\mathbf{A}_1(\theta_1,\varphi_1)\widetilde{\mathbf{x}}_1+\alpha\xi_2\mathbf{A}_2(\theta_2,\varphi_2)\widetilde{\mathbf{x}}_2$ and $\mathbf{u}_2=\alpha\xi_3\mathbf{A}_2(\theta_2,\varphi_2)\widetilde{\mathbf{x}}_2+\alpha\xi_2\mathbf{A}_1(\theta_1,\varphi_1)\widetilde{\mathbf{x}}_1$. In fact, we are more concerned about the position of the target rather than the reflection coefficient. Thus, the CRB matrix for estimating $\boldsymbol{\rho}_k$ are given by
\begin{equation}
	\begin{aligned}\label{46}
		\mathbf{F}_{\boldsymbol{\rho}_k} = \mathbf{F}_{\boldsymbol{\rho}_k \boldsymbol{\rho}_k}-\mathbf{F}_{\boldsymbol{\rho}_k \alpha}F_{\alpha \alpha}^{-1}F_{\boldsymbol{\rho}_k \alpha}^T.
	\end{aligned}
\end{equation}
According to (\ref{Fij}), the partial derivations of $\mathbf{u}_k$ with respect to location parameters can be calculated as
\begin{equation}
	\begin{aligned}
		\mathbf{F}_{\boldsymbol{\rho}_k \boldsymbol{\rho}_k} = \frac{2}{\sigma_k^2} \text{Re} \bigg\{ \frac{\partial \mathbf{u}_k^H}{\partial \boldsymbol{\rho}_k}  \frac{\partial \mathbf{u}_k}{\partial \boldsymbol{\rho}_k}   \bigg\},
	\end{aligned}
\end{equation}

\begin{equation}
	\begin{aligned}
		\mathbf{F}_{\boldsymbol{\rho}_k \alpha} = \frac{2}{\sigma_k^2} \text{Re} \bigg\{ \frac{\partial \mathbf{u}_k^H}{\partial \boldsymbol{\rho}_k}  \frac{\partial \mathbf{u}_k}{\partial \alpha}   \bigg\},
	\end{aligned}
\end{equation}
and
\begin{equation}
	\begin{aligned}
		\mathbf{F}_{\alpha \alpha} = \frac{2}{\sigma_k^2} \text{Re} \bigg\{ \frac{\partial \mathbf{u}_k^H}{\partial \alpha}  \frac{\partial \mathbf{u}_k}{\partial \alpha}   \bigg\},
	\end{aligned}
\end{equation}
where
\begin{equation}
	\begin{aligned}
		\frac{\partial \mathbf{u}_1}{\partial \boldsymbol{\rho}_1}=\alpha \xi_1 \bigg[  \frac{  \partial \mathbf{A}_1(\theta_1,\varphi_1)}{\partial \theta_1} \widetilde{\mathbf{x}}_1, \frac{  \partial \mathbf{A}_1(\theta_1,\varphi_1)}{\partial \varphi_1} \widetilde{\mathbf{x}}_1, \mathbf{A}_1 \frac{\partial \tilde{\mathbf{x}}_1}{\partial \tau_1}    \bigg],
	\end{aligned}
\end{equation}

\begin{equation}
	\begin{aligned}
		\frac{\partial \mathbf{u}_2}{\partial \boldsymbol{\rho}_2}=\alpha \xi_3 \bigg[ \frac{  \partial \mathbf{A}_2(\theta_2,\varphi_2)}{\partial \theta_2}, \frac{  \partial \mathbf{A}_2(\theta_2,\varphi_2)}{\partial \varphi_2},\mathbf{A}_2\frac{\partial \widetilde{\mathbf{x}}_2}{\partial \tau_2}   \bigg] \widetilde{\mathbf{x}}_2,
	\end{aligned}
\end{equation}

\begin{equation}
	\begin{aligned}
		\frac{\partial \mathbf{u}_1}{\partial \alpha}=\xi_1\mathbf{A}_1(\theta_1,\varphi_1)\widetilde{\mathbf{x}}_1+\xi_2\mathbf{A}_2(\theta_2,\varphi_2)\widetilde{\mathbf{x}}_2,
	\end{aligned}
\end{equation}
and
\begin{equation}\label{47}
	\begin{aligned}
		\frac{\partial \mathbf{u}_2}{\partial \alpha}=\xi_3\mathbf{A}_2(\theta_2,\varphi_2)\widetilde{\mathbf{x}}_2+\xi_2\mathbf{A}_1(\theta_1,\varphi_1)\widetilde{\mathbf{x}}_1.
	\end{aligned}
\end{equation}
By substituting equations (\ref{46})-(\ref{47}) into equations (\ref{C1})-(\ref{C2}), we can obtain the CRB matrix with closed-form expressions to evaluate the sensing performance.

From the derivation mentioned above, we can find that both communication and sensing performance are related with the beamforming vector $\mathbf{w}_i$ and $\mathbf{v}_m$. Therefore, in the next section, we will design beamforming at the TBS and the LEO satellite to improve the communication and sensing performance of ISTMS.

\section{Beamforming Design}
In this section, we focus on jointly optimizing the TBS's and the LEO satellite's beamforming to improve the communication and sensing performance. Specifically, we propose a joint beamforming design algorithm to maximize the sum data transmission rate of maritime users while satisfying target localization accuracy requirements and transmit power constraints.

\subsection{Problem Formulation}
To enhance the overall performance of the proposed ISTMS, we aim to maximize the sum data transmission rate of maritime users while satisfying target localization accuracy requirements and transmit power constraints. Specifically, the optimization problem can be formulated as
\begin{subequations}
	\begin{eqnarray}
\!\!\!\!\!\!\!\!\!\!\!\!\!\!\!\!\!\!\!\!\!\!\!\!\!\!\!\!\!		\mathcal{Q}1:	\underset{\mathbf{w}_i,\mathbf{v}_m}{\mathop{\text{max}}}\,&&\sum_{i=1}^{K_1}R_{1,i}+\sum_{m=1}^{K_2}R_{2,m} \label{OP1obj}\\
		\textrm{s.t.}&& \text{tr}(\mathbf{C}_1) \leq \varepsilon_1,\label{OP1st1}\\
		&& \text{tr}(\mathbf{C}_2) \leq \varepsilon_2,\label{OP1st2}\\
		&& P_T\leq P_1^{\max},\label{OP1st3}\\
		&& P_S\leq P_2^{\max}, \label{OP1st4}
	\end{eqnarray}
\end{subequations}
where $P_1^{\max}$ and $P_2^{\max}$ represent the maximum transmit power budget of the TBS and the LEO satellite, respectively. $\varepsilon_1$ and $\varepsilon_2$ are the minimum required location accuracy of the TBS and the LEO satellite, respectively. Notice that the both beamforming vectors $\mathbf{w}_i$ and $\mathbf{v}_m$ are highly coupled in the objective and constraint functions. To this end, based on semi-definite relaxation (SDR), we can define $\mathbf{W}_i=\mathbf{w}_i\mathbf{w}_i^H$ and $\mathbf{V}_m=\mathbf{v}_m\mathbf{v}_m^H$. Obviously, the constraints (\ref{OP1st1}) and (\ref{OP1st2}) are still non-convex. To solve this problem, we introduce the auxiliary matrix $\mathbf{U}_1 \in \mathbb{R}^{3\times 3}$ and $\mathbf{U}_2 \in \mathbb{R}^{3\times 3}$. Then, we substitute equations (\ref{C1}) and (\ref{C2}) into the constraints (\ref{OP1st1}) and (\ref{OP1st2}), respectively, which can be given as
\begin{equation}
	\begin{aligned}
		\mathbf{C}_1=\big[\mathbf{J}_1^T(\mathbf{F}_{\boldsymbol{\rho}_1 \boldsymbol{\rho}_1}-\mathbf{F}_{\boldsymbol{\rho}_1 \alpha}F_{\alpha \alpha}^{-1}F_{\boldsymbol{\rho}_1 \alpha}^T)\mathbf{J}_1\big]^{-1} \succeq \mathbf{U}_1,
	\end{aligned}
\end{equation}
and
\begin{equation}
	\begin{aligned}
		\mathbf{C}_2=\big[\mathbf{J}_2^T(\mathbf{F}_{\boldsymbol{\rho}_2 \boldsymbol{\rho}_2}-\mathbf{F}_{\boldsymbol{\rho}_2 \alpha}F_{\alpha \alpha}^{-1}F_{\boldsymbol{\rho}_2 \alpha}^T)\mathbf{J}_2\big]^{-1} \succeq \mathbf{U}_2.
	\end{aligned}
\end{equation}
According to the Schur complement theorem, the constraints (\ref{OP1st1}) and (\ref{OP1st2}) can be further transformed into the convex form, which can be expressed as
\begin{equation} \label{jz1}
	\begin{aligned}
        \begin{bmatrix}
        	\mathbf{J}_1\mathbf{F}_{\boldsymbol{\rho}_1 \boldsymbol{\rho}_1}\mathbf{J}_1-\mathbf{U}_1 & \mathbf{J}_1^T\mathbf{F}_{\boldsymbol{\rho}_1\alpha}\\
            \mathbf{F}_{\boldsymbol{\rho}_1 \alpha}^T\mathbf{J}_1  & \mathbf{F}_{\alpha \alpha}
        \end{bmatrix} \succeq 0,
	\end{aligned}
\end{equation}
and
\begin{equation}  \label{jz2}
	\begin{aligned}
		\begin{bmatrix}
			\mathbf{J}_2\mathbf{F}_{\boldsymbol{\rho}_2 \boldsymbol{\rho}_2}\mathbf{J}_2-\mathbf{U}_2 & \mathbf{J}_2^T\mathbf{F}_{\boldsymbol{\rho}_2\alpha}\\
			\mathbf{F}_{\boldsymbol{\rho}_2 \alpha}^T\mathbf{J}_2  & \mathbf{F}_{\alpha \alpha}
		\end{bmatrix} \succeq 0,
	\end{aligned}
\end{equation}
where constraints (\ref{jz1}) and (\ref{jz2}) can be regarded as the standard linear matrix inequality, which is convex and convenient to solve directly. Thus, the optimization problem can be rewritten as
\begin{subequations}
	\begin{eqnarray}
	\!\!\!\!\!\!\!\!\!\!\!\!\!\!\!\!			\mathcal{Q}2:	\underset{\mathbf{w}_i,\mathbf{v}_m}{\mathop{\text{max}}}\,&&\sum_{i=1}^{K_1}R_{1,i}+\sum_{m=1}^{K_2}R_{2,m} \label{OP2obj}\\
		\textrm{s.t.}&& \nonumber (\ref{jz1}),(\ref{jz2}),\\ 	
		&& \text{tr}(\mathbf{U}_1) \leq \varepsilon_1, \quad \mathbf{U}_1 \succeq 0,\label{OP2st1}\\
		&& \text{tr}(\mathbf{U}_2) \leq \varepsilon_2, \quad \mathbf{U}_2 \succeq 0, \label{OP2st2}\\
		&& \text{tr}\bigg(\sum_{i=1}^{K_1} \mathbf{W}_i\bigg)\leq P_1^{\max},\label{OP2st3}\\
		&& \text{tr}\bigg(\sum_{m=1}^{K_2} \mathbf{V}_m \bigg)\leq P_2^{\max}, \label{OP2st4}\\
		&& \text{rank}(\mathbf{W}_i) =1, \mathbf{W}_i \succeq0, \forall i, \label{OP2st5}\\
		&& \text{rank}(\mathbf{V}_m) =1, \mathbf{V}_m \succeq0, \forall m. \label{OP2st6}
	\end{eqnarray}
\end{subequations}

However, the objective function in problem $\mathcal{Q}2$ is still convex. Therefore, we use the successive convex approximation method to deal with it, which approximates the non-convex objective function as a convex function in the iterations. Then, by introducing new auxiliary matrix $\mathbf{H}_{1,i}=\mathbf{h}_{1,i}\mathbf{h}_{1,i}^H$ and $\mathbf{H}_{2,m}=\mathbf{h}_{2,m}\mathbf{h}_{2,m}^H$, and substituting (\ref{SINR1}) and (\ref{SINR2}) into (\ref{R1}) and (\ref{R2}), respectively, the objective function can be rewritten as
\begin{equation}  \label{zhankai}
	\begin{aligned}
		&R_{1,i}=\\
		&\log_2\bigg(1+\frac{\text{tr}(\mathbf{H}_{1,i}\mathbf{W}_i)}{\sum_{j=1,j\ne   i}^{K_1}\text{tr}(\mathbf{H}_{1,i}\mathbf{W}_j)+\sum_{m=1}^{K_2}\text{tr}(\mathbf{H}_{2,i}\mathbf{V}_m)+\sigma_{1,i}^{2}}\bigg)\\
		&=\log_2\bigg(\sum_{j=1}^{K_1}\text{tr}(\mathbf{H}_{1,i}\mathbf{W}_j)+\sum_{m=1}^{K_2}\text{tr}(\mathbf{H}_{2,i}\mathbf{V}_m)+\sigma_{1,i}^{2}\bigg)-\\
		&\quad \quad \log_2\bigg(\sum_{j=1,j\ne i}^{K_1}\text{tr}(\mathbf{H}_{1,i}\mathbf{W}_j)+\sum_{m=1}^{K_2}\text{tr}(\mathbf{H}_{2,i}\mathbf{V}_m)+\sigma_{1,i}^{2}\bigg).
	\end{aligned}
\end{equation}
By observing (\ref{zhankai}), we can see that the first term is a non-concave function, and the second term is a convex function, resulting in the form of non-concave plus concave. Fortunately, the first term can be expanded in the form of concave minus concave. Thus, we only need to perform the first-order Taylor expansion on the second term, which can be expressed as
\begin{equation}
	\begin{aligned}
		&R_{1,i}\approx\\
		&\quad \log_2\bigg(\sum_{j=1}^{K_1}\text{tr}(\mathbf{H}_{1,i}\mathbf{W}_j)+\sum_{m=1}^{K_2}\text{tr}(\mathbf{H}_{2,i}\mathbf{V}_m)+\sigma_{1,i}^{2}\bigg)-\\
		&\quad \quad \log_2\big(A_{1,i}\big)-\sum_{j=1,j\neq i}^{K_1}\text{tr}\big(B_{1,i} (\mathbf{W}_i-\mathbf{W}_i^{'})  \big)-\\
		&\quad \quad \quad \quad  \sum_{m=1}^{K_2}\text{tr}\big(B_{1,i} (\mathbf{V}_m-\mathbf{V}_m^{'}) \triangleq \tilde{R}_{1,i},
	\end{aligned}
\end{equation}
where $\mathbf{W}_i^{'}$ and $\mathbf{V}_m^{'}$ are obtained by using the successive convex approximation (SCA) method \footnote{SCA typically converges to a stationary point, which may be a local optimum or a saddle point, rather than the global solution \cite{R2}}. $A_{1,i}=\sum_{j=1,j\ne i}^{K_1}\text{tr}(\mathbf{H}_{1,i}\mathbf{W}_j^{'})+\sum_{m=1}^{K_2}\text{tr}(\mathbf{H}_{2,i}\mathbf{V}_m^{'})+\sigma_{1,i}^{2}$ and $B_{1,i}=\log_2(e)\frac{\mathbf{H}_{1,i}}{A_{1,i}}$ are the new introduced variable. Similarly, we can approximate $R_{2,m}$ as
\begin{equation}
	\begin{aligned}
		&R_{2,m}\approx\\
		&\quad \log_2\bigg(\sum_{k=1}^{K_2}\text{tr}(\mathbf{H}_{2,m}\mathbf{V}_k)+\sigma_{2,m}^{2}\bigg)-\log_2\big(A_{2,m}\big)-\\
		&\quad \quad \quad \quad \sum_{k=1,k\neq m}^{K_2}\text{tr}\big(B_{2,m} (\mathbf{V}_m-\mathbf{V}_m^{'})\triangleq \tilde{R}_{2,m},
	\end{aligned}
\end{equation}
where $A_{2,m}=\sum_{k=1,k\ne m}^{K_2}\text{tr}(\mathbf{H}_{2,m}\mathbf{V}_k^{'})+\sigma_{2,m}^{2}$ and $B_{2,m}=\log_2(e)\frac{\mathbf{H}_{2,m}}{A_{2,m}}$. By using the difference of convex algorithm approach, it is evident that the objective function is already convex. Thus, the optimization problem can be reformulated as
\begin{subequations}
	\begin{eqnarray}
		\!\!\!\!\!\!\!\!\!\!\!\!\!\!\!\!\!\!\!\!		\mathcal{Q}3:	\underset{\mathbf{w}_i,\mathbf{v}_m}{\mathop{\text{max}}}\,&&\sum_{i=1}^{K_1}\tilde{R}_{1,i}+\sum_{m=1}^{K_2}\tilde{R}_{2,m} \label{OP3obj}\\
		\textrm{s.t.}&& \nonumber (\ref{jz1}),(\ref{jz2}),\\ 	
		&& \nonumber (\ref{OP2st1})-(\ref{OP2st6}).
	\end{eqnarray}
\end{subequations}

\begin{figure*}[ht] 
	\begin{subequations}
		\begin{eqnarray}
			\mathcal{Q}4:	\underset{\mathbf{W}_i,\mathbf{V}_m}{\mathop{\text{max}}}\,\!\!\!\!\!&&\!\!\!\!\!\sum_{i=1}^{K_1}\tilde{R}_{1,i}+\sum_{m=1}^{K_2}\tilde{R}_{2,m}+\delta \bigg(\sum_{i=1}^{K_1}\big((\tilde{\mathbf{e}}_{1,i}^{(t)})^H\mathbf{W}_i^{(t+1)}\tilde{\mathbf{e}}_{1,i}^{(t)}+\sum_{m=1}^{K_2}\big((\tilde{\mathbf{e}}_{2,m}^{(t)})^H\mathbf{V}_m^{(t+1)}\tilde{\mathbf{e}}_{2,m}^{(t)}\big)\bigg) \label{OP4obj}\\
			\textrm{s.t.}&&  \nonumber (\ref{jz1}),(\ref{jz2}),\\ 	
			&& \nonumber (\ref{OP2st1})-(\ref{OP2st4}),\\
			&& \mathbf{W}_i\succeq 0,\forall i,\label{OP4st1}\\
			&& \mathbf{V}_m\succeq 0,\forall m\label{OP4st2}
		\end{eqnarray}
	\end{subequations}
		\centering
	\rule[-10pt]{18cm}{0.07em}
\end{figure*}

Yet, the rank-one constraints (\ref{OP2st5}) and (\ref{OP2st6}) are still non-convex due to the SDR technique. Herein, we propose to use a penalty function method to eliminate the rank-one constraints, which can make the optimization problem convex. Since $\mathbf{W}_i$ and $\mathbf{V}_m$ are all positive semi-definite matrices, we can use the following equations to replace the rank-one constraints as \cite{st4}
\begin{equation}
	\begin{aligned}
		\tilde{\lambda}_{1,i}=\text{tr}(\mathbf{W}_i),
	\end{aligned}
\end{equation}
and
\begin{equation}
	\begin{aligned}
		\tilde{\lambda}_{2,m}=\text{tr}(\mathbf{V}_m),
	\end{aligned}
\end{equation}
where the $\tilde{\lambda}_{1,i}$ and $\tilde{\lambda}_{2,m}$ represent the maximum eigenvalue of the $\mathbf{W}_i$ and $\mathbf{V}_m$, respectively. By introducing the penalty function, the objective function of problem $\mathcal{Q}3$ can be rewritten as
\begin{equation}  \label{new}
	\begin{aligned}
		&\underset{\mathbf{w}_i,\mathbf{v}_m}{\mathop{\text{max}}}\,\sum_{i=1}^{K_1}\tilde{R}_{1,i}+\sum_{m=1}^{K_2}\tilde{R}_{2,m}+\\
		&\quad \quad \quad \delta\bigg(\sum_{i=1}^{K_1}\big(\mathbf{W}_i-\lambda_{1,i}^{\max}\big)+\sum_{m=1}^{K_2}\big(\mathbf{V}_m-\lambda_{2,m}^{\max}\big)\bigg),
	\end{aligned}
\end{equation}
where $\delta > 0$ denotes the penalty factor. Due to the presence of the penalty function, the objective function (\ref{new}) is still non-convex. To deal with it, we use SCA to obtain the $\mathbf{W}_i^{(t)}$ and $\mathbf{V}_m^{(t)}$ during the $t$-th iteration. Specifically, we can obtain the inequalities as
\begin{equation}
	\begin{aligned}
		&\text{tr}(\mathbf{W}_i^{(t+1)})-(\tilde{\mathbf{e}}_{1,i}^{(t)})^H\mathbf{W}_i^{(t+1)}\tilde{\mathbf{e}}_{1,i}^{(t)} \\ &\quad \quad \quad  \text{tr}(\mathbf{W}_i^{(t+1)})-\tilde{\lambda}_{1,i}^{(t+1)}\geq 0,
	\end{aligned}
\end{equation}
and
\begin{equation}
	\begin{aligned}
		&\text{tr}(\mathbf{V}_m^{(t+1)})-(\tilde{\mathbf{e}}_{w,m}^{(t)})^H\mathbf{V}_m^{(t+1)}\tilde{\mathbf{e}}_{w,m}^{(t)} \\ &\quad \quad \quad  \text{tr}(\mathbf{V}_m^{(t+1)})-\tilde{\lambda}_{w,m}^{(t+1)}\geq 0,
	\end{aligned}
\end{equation}
where $\tilde{\mathbf{e}}_{1,i}$ and $\tilde{\mathbf{e}}_{2,m}$ are the unit eigenvector of the $\tilde{\lambda}_{1,i}$ and $\tilde{\lambda}_{2,m}$, respectively. With this upper bound, we can transform the optimization problem into convex. Finally, the optimization problem $\mathcal{Q}4$ is shown at top of the page, which can be solved directly. The proposed beamforming design for joint communication and sensing in ISTMS is summarized in Algorithm 2. {It is worth noting that to ensure system robustness, a communication-priority fallback strategy is adopted when the joint optimization problem becomes infeasible. Specifically, the system primarily focuses on maximizing the sum data rate of maritime users, and subsequently allocates the remaining power budget to the sensing function.}

\begin{algorithm}[!h]
	\caption{Beamforming Design for Joint Communication and Sensing in ISTMS}
	\label{alg:AOA}
	\renewcommand{\algorithmicrequire}{\textbf{Input:}}
	\renewcommand{\algorithmicensure}{\textbf{Output:}}
	\begin{algorithmic}[1]
		\REQUIRE $M_1$, $M_2$, $K_1$, $K_2$, $P_1^{\max}$, $P_2^{\max}$, $\varepsilon_1$, $\varepsilon_2$, $\sigma^2_0$, $\delta$  
		\ENSURE $\mathbf{v}_m$, $\mathbf{w}_i$   
		\STATE $\mathbf{Initialization:}$ Set accuracy $\epsilon_1$, $\epsilon_2$, maximal iteration number $t_{\max}$, penalty factor $\delta$, positive constant $\tau$, initial feasible solution $\mathbf{V}_m^{(0)}$, $\mathbf{W}_i^{(0)}$ and iteration index $t=0$
		\STATE $\mathbf{repeat}$
		\STATE Update $\mathbf{W}_i^{'}=\mathbf{W}_i^{(t)}$, $\mathbf{V}_m^{'}=\mathbf{V}_m^{(t)}$
		\STATE \ \  Solving problem ($\mathcal{Q}4$) to get $\mathbf{W}_i^{(t+1)}$ and $\mathbf{V}_m^{(t+1)}$
		\IF {$|\text{tr}(\mathbf{W}_i^{(t+1)})-\tilde{\lambda}_{1,i}^{(t+1)}|>\epsilon_1$ and \\
			\quad \quad \quad \quad $|\text{tr}(\mathbf{V}_m^{(t+1)})-\tilde{\lambda}_{2,m}^{(t+1)}|>\epsilon_2$}
		\STATE Set $\delta=\delta+\tau$
		\ENDIF
		\STATE Set $t=t+1$
		\STATE $\mathbf{until}$ $t=t_{\max}$ or convergence
		\STATE Finally, apply eigenvalue decomposition (EVD) to $\mathbf{W}_i^{(t+1)}$, $\mathbf{V}_m^{(t+1)}$ and get $\mathbf{w}_i$, $\mathbf{v}_m$.
	\end{algorithmic}
\end{algorithm}

\subsection{Algorithm Analysis}
In this subsection, we analyze the performance of the proposed algorithms in terms of computational complexity and convergence behavior.

\emph{Convergence Analysis:} The proposed Algorithm 1 is a heuristic algorithm based on the DE method. The convergence of Algorithm 1 refers to find the global optimal solution within a finite time, which usually comes from experimental research and experience. To visualize the convergence of proposed Algorithm 1, we simulate the convergence of Algorithm 1 under different population sizes in Fig. $\ref{diedai1}$. {Moreover, we provide the theoretical analysis of the convergence of the DE method in Appendix A \cite{cv1}.} Furthermore, for the proposed Algorithm 2, the objective function and constraints are all convex, which can ensure to obtain the solution during limited iterations. Based on Algorithm 2, we find that the solutions $\mathbf{W}_i^{(t+1)}$ and $\mathbf{V}_m^{(t+1)}$ of $\mathcal{Q}4$ are always less than or equal to the solutions $\mathbf{W}_i^{(t)}$ and $\mathbf{V}_m^{(t)}$. Due to the power and location sensing constraints, the problem $\mathcal{Q}4$ has a upper bound. According to the monotonic bounded criterion, it indicates that Algorithm 2 can converge to a stationary value within iterations.

\emph{Complexity Analysis:} The computational complexity of the location sensing Algorithm 1 mainly depends on the population size, the maximum number of iterations and the dimension of problem. In each iteration, the Algorithm 1 evaluates the fitness value of each individual and updates the population through cross and mutation. Therefore, the computational complexity of Algorithm 1 is $\mathcal{O}(N_pG_{\max}d_{\mathrm{rand}})$. The optimization problem ($\mathcal{Q}4$) consists of linear matrix inequality (LMI) constraints. Therefore, following the conventional traditional standard interior-point method (IPM) \cite{complexity}, the computational cost of Algorithm 2 depends on two factors: the total iterations required and complexity per iteration. For a given precision $\eta >0$, the iteration count needed to reach an $\eta$-optimal solution for problem ($\mathcal{Q}4$) scales roughly as $\zeta \cdot \ln(1/\eta)$, where $\zeta$ denotes the barrier parameter quantifying the geometric complexity of the conic constraints. Consequently, the total computational complexity of Algorithm 2 is $\zeta \cdot \ln(1/\eta)$ with $\zeta=\sqrt{M_1K_1+M_2K_2+M_1+M_2+8}\cdot [n_1 \cdot (K_1M_1^3+K_1+64)+n_2 \cdot (K_2M_2^3+K_2+64)]$, where $n_1=\mathcal{O}(K_1M_1^2)$ and $n_2=\mathcal{O}(K_2M_2^2)$.

{\subsection{Implementation Complexity and Hardware Requirements}
\emph{1. Implementation Complexity:}
Algorithm~1 is based on a standard DE framework, which is easy to implement with simple control flow and a small number of tunable parameters. It requires only basic arithmetic operations and fitness evaluations, and its evolution process can be parallelized across the population. The algorithm typically converges within a few tens of generations, making it suitable for real-time operation without heavy computational burden. Algorithm~2 employs successive convex approximation to solve a semidefinite programming relaxation. Although this involves an iterative procedure with convex subproblems, the overall structure remains modular and can be readily implemented using existing convex optimization solvers or custom code for linear matrix inequalities. The penalty method used to enforce rank-one constraints adds a tunable penalty factor but does not introduce algorithmic complexity beyond standard SCA.

\emph{2. Hardware Requirements:}
Practical deployment of the proposed system requires a terrestrial base station equipped with a uniform planar array of $M_1$ antennas  operating at VHF band (160~MHz), a total transmit power up to 47~dBm, and a high-speed feeder link to the LEO satellite. The LEO satellite payload includes a uniform planar array of $M_2$ antennas, a maximum antenna gain of 55~dBi, and a radiation-hardened DSP for beamforming computation. The signal processing unit can be a low-cost microcontroller for Algorithm~1, whereas Algorithm~2 demands a more powerful processor such as an ARM Cortex-A or an FPGA. A GNSS receiver with pulse-per-second output provides coarse synchronization; a chip-scale atomic clock may be added for high-precision coherent processing. All these components are within the current state of the art for maritime communication infrastructure \cite{R3}, \cite{R4}.}

\section{Simulation Results}
In this section, we present extensive numerical simulations to validate the performance of proposed algorithms. The primary simulation parameters are detailed in Table \uppercase\expandafter{\romannumeral2}{ \footnote{Note that a 15 MHz bandwidth is adopted to ensure sufficient range resolution for sensing, while the satellite antenna gain is set to 55 dBi to represent the aggregate peak gain, which ensures link closure for high-precision maritime ISAC systems.}}.

\begin{table}[h]
	\small
	\centering
	\caption{Primary Simulation Parameters of ISTMS}\label{Simulation}
	\begin{tabular}{|c|c|}
		\hline
		Parameter & Value\\\hline \hline
		Initial cross factor   & 0.5 \\\hline
		Population size          &   30 \\\hline
		Population dimension   & 3  \\\hline
		Satellite orbit & LEO \\\hline
		Altitude of orbit      &  550 km  \\\hline
		Effective antennas height of users  & 10m \\\hline
		Effective antennas height of TBS   & 50m \\\hline
		3dB angle & $0.4^{\circ}$ \\\hline
		Rician factor  & 10 \\\hline
		Rain fading variance  &  1.63dB \\\hline
		Rain fading mean &  -2.6dB\\\hline
		Boltzmann's constant  &  1.38$\times 10^{-23}$ J/m\\ \hline
		Variance of AWGN & -110dBm\\ \hline
		TBS total transmit power   & 47dBm \\ \hline
		Carrier frequency &  160 MHz \\\hline
		Bandwidth  &  15 MHz \\\hline
		Satellite antenna gain   &  55 dBi \\\hline
		TBS antenna gain   &  30 dBi \\\hline
		Receive antenna gain  & 20 dBi \\\hline
		Noise temperature   &  300 K \\\hline	
	\end{tabular}
\end{table}	

\begin{figure}[h]
	\centering
	\includegraphics[width=0.5\textwidth]{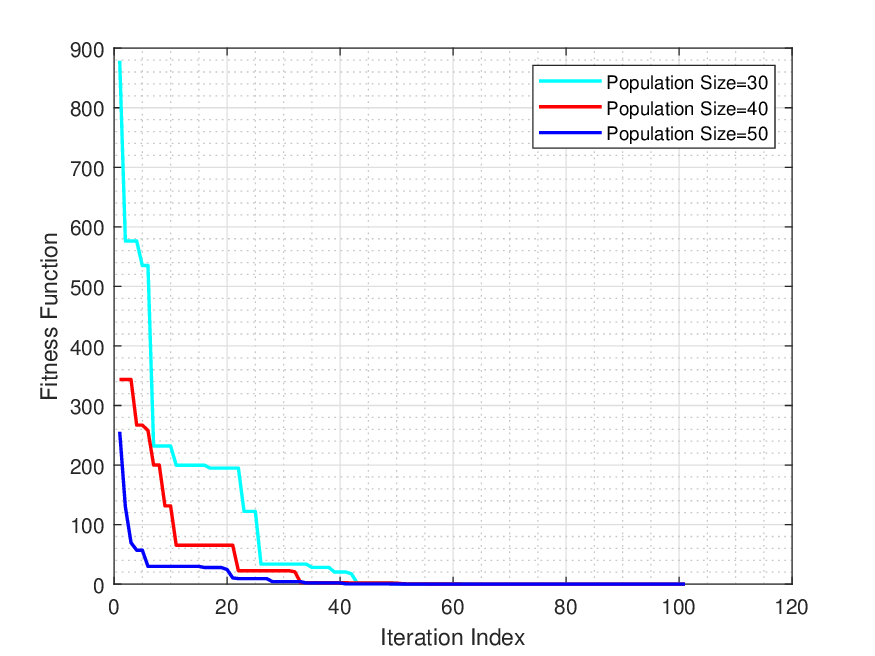} 
	\caption{Convergence performance of the proposed Algorithm 1 with different population size.}
	\label{diedai1}
\end{figure}

\begin{figure}[h]
	\centering
	\includegraphics[width=0.5\textwidth]{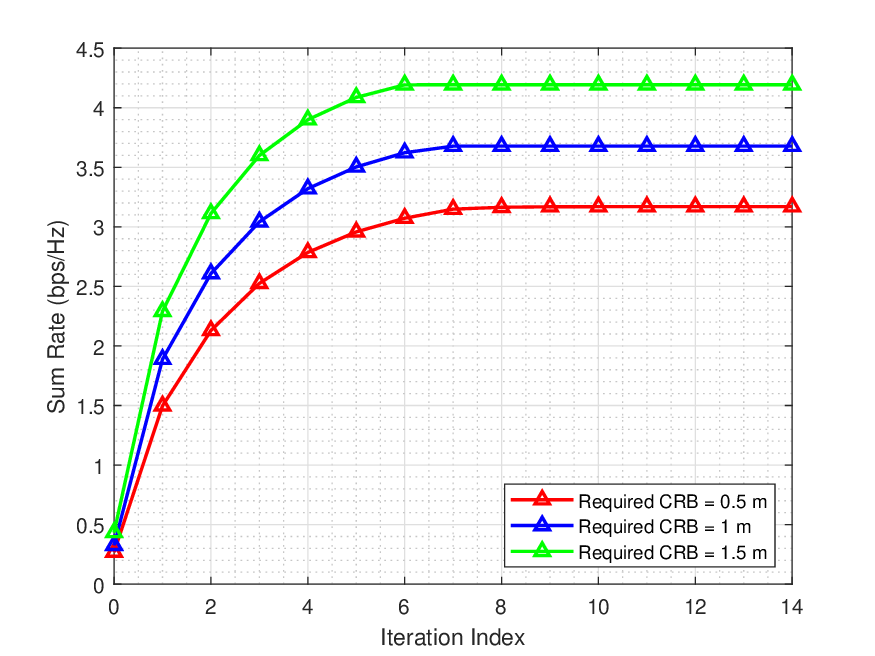} 
	\caption{Convergence performance of the proposed Algorithm 2 with different required location accuracy.}
	\label{diedai2}
\end{figure}

We first analyze the convergence performance of Algorithms 1 and 2. As for Algorithm 1, convergence is assessed through the evolution of its fitness function. As shown in Fig. $\ref{diedai1}$, the fitness function decreases monotonically, achieving convergence within 40 iterations, which confirms the rapid convergence characteristics of proposed Algorithm 1. Furthermore, increasing the population size enhances both population diversity and the algorithm's global search capability. Meanwhile, the expanded search space facilitates to escape from local optima while promoting faster convergence to the global optimum.

Then, Fig. $\ref{diedai2}$ illustrates the convergence behavior of Algorithm 2 under varying CRB requirements. As shown in the Fig. $\ref{diedai2}$, the sum rate gradually converges to a stable value within 15 iterations, which proves the effectiveness of the proposed Algorithm 2. The result shows that the sum rate of the maritime users follows a characteristic growth trajectory with iterations before stabilizing. Notably, the achievable sum rate of maritime users exhibits an inverse relationship with CRB requirements. This is because that the higher localization demands greater allocation of power and computational resources, which leads to a decline in the data transmission rate of maritime users.

\begin{figure}[h]
	\centering
	\includegraphics[width=0.5\textwidth]{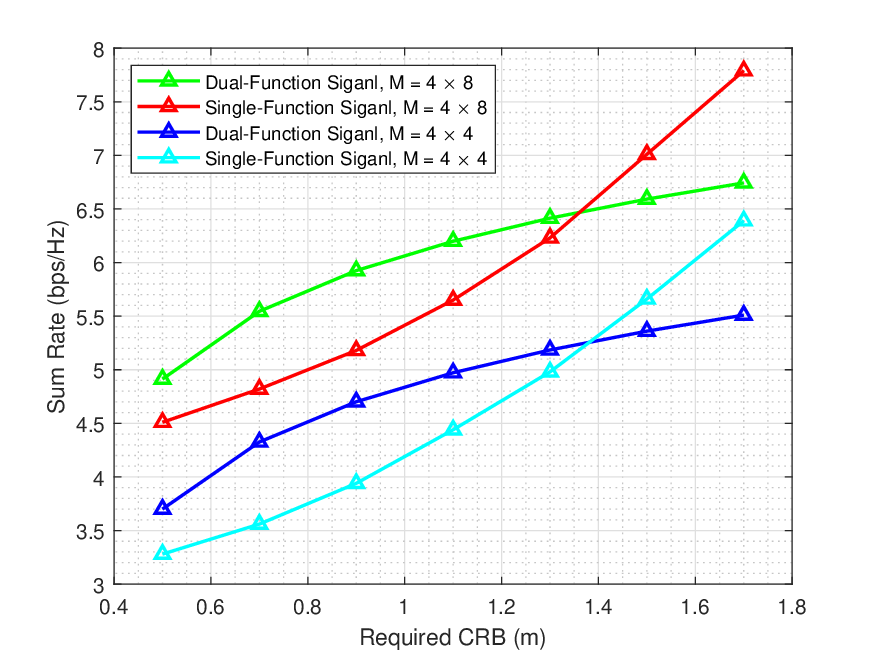} 
	\caption{Sum rate versus required location accuracy with different signal models and number of antennas.}
	\label{signal}
\end{figure}

Furthermore, we investigate the impact of antenna array size and transmit signal structures. Two distinct signal frameworks are considered. The dual-function signal simultaneously performs communication and sensing by the same signals. The single-function signal is designed by employing dedicated signals for each functionality. As demonstrated in Fig. $\ref{signal}$, the sum rate of maritime users exhibits consistent improvement with increasing antenna array size across all localization accuracy requirements. Because of the increasing number of deployed antennas, there is more available spatial degrees of freedom to design, thereby enabling higher array gain potential. In addition, although the single-function signal can have an additional design for sensing beamforming, the sensing signals also cause serious interference to the communication signal, resulting in a reduction in communication performance. Our results reveal an inflection point where reduced localization accuracy requirements correspond to diminished interference levels from sensing waveforms. Therefore, in the practical applications, it is necessary to design appropriate transmit signals and antenna scales for the TBS and the LEO satellite based on different communication and sensing requirements.

\begin{figure}[h]
	\centering
	\includegraphics[width=0.5\textwidth]{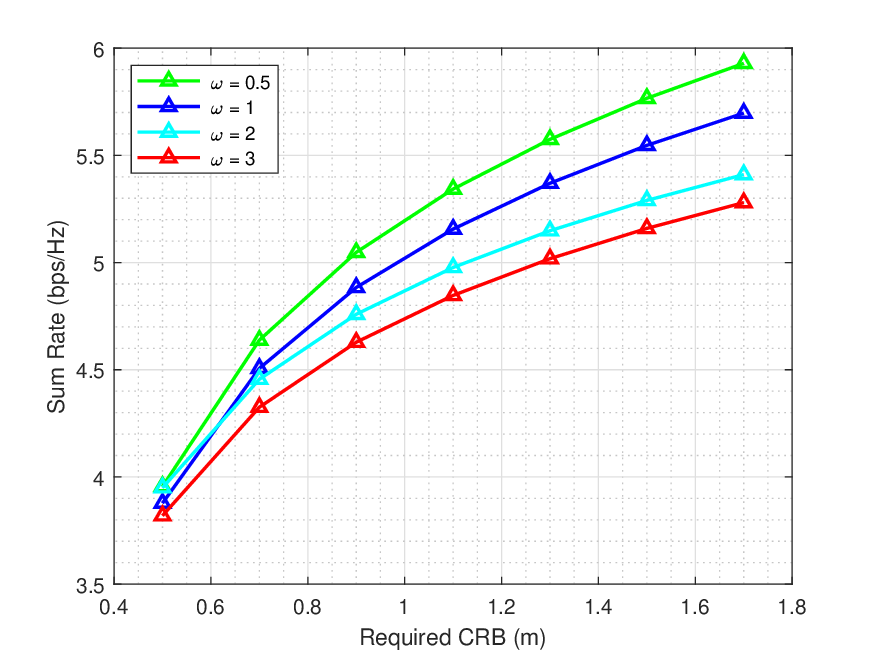} 
	\caption{Sum rate versus required location accuracy with different ratios of NSUs to OSUs.}
	\label{midu}
\end{figure}

Next, we examine the effect of maritime user density. we define the maritime user density as $\omega=K_1/K_2$. Fig. $\ref{midu}$ reveals an inverse relationship between user density and sum rate. This degradation stems from increased co-channel interference in near-shore regions, where users experience overlapping coverage from both the TBS and the LEO satellite. Therefore, when the maritime user density increases, the sum rate of maritime users decreases. The simulation result highlights a critical design trade-off between network capacity and user density in maritime communication systems, particularly relevant for crowded near-shore areas and busy shipping lanes.

\begin{figure}[h]
	\centering
	\includegraphics[width=0.5\textwidth]{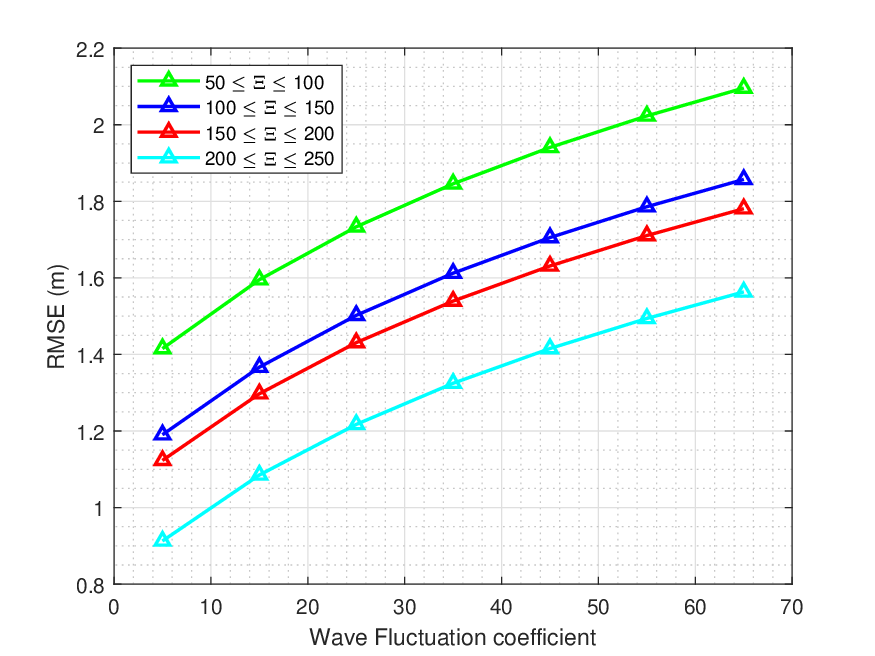} 
	\caption{RMSE versus wave fluctuation coefficient with different location relationship between target and maritime users.}
	\label{case}
\end{figure}

Then, we reveal the performance of joint communication and sensing in different positional cases of target. We choose to use the root of mean square error (RMSE) between the estimated target location obtained through Algorithm 1 and the its true position, which can be expressed as $\mathrm{RMSE}=\sqrt{\frac{1}{100}\sum_{i=1}^{100}||\mathbf{p}_i^*-\mathbf{p}_i||^2}$. We denote $[x_i,y_i,z_i]$ as the location vector of the $i$-th maritime user. Further, we define the location relationship between the target and the maritime users as $\Xi=\sqrt{\frac{1}{K_1+K_2}\sum_{i=1}^{K_1+K_2}(p_x-x_i)^2+(p_y-y_i)^2+(p_z-z_i)^2}$. When the wave fluctuations are violent, the height difference between the target and the base station will also increase, so the $p_z$ needs to be set larger. On the contrary, when the wave fluctuations are calm, the $\mathbf{p}$ are set with a smaller $p_z$. To assess the degree of wave fluctuations, we define wave fluctuation coefficient as $\sqrt{\frac{1}{100}\sum_{i=1}^{100}(p_{z,i}-x_t)^2}$. From Fig. $\ref{case}$, we can see that when the target is separated from the location of users, the interference to the sensing echo signal is reduced, and the location accuracy is also improved. Moreover, when the wave fluctuations are violent, the target direction is relatively farther from the TBS. Due to the worse direction matching degree, the transmit signal is difficult to be effectively used for location sensing.

\begin{figure}[h]
	\centering
	\includegraphics[width=0.5\textwidth]{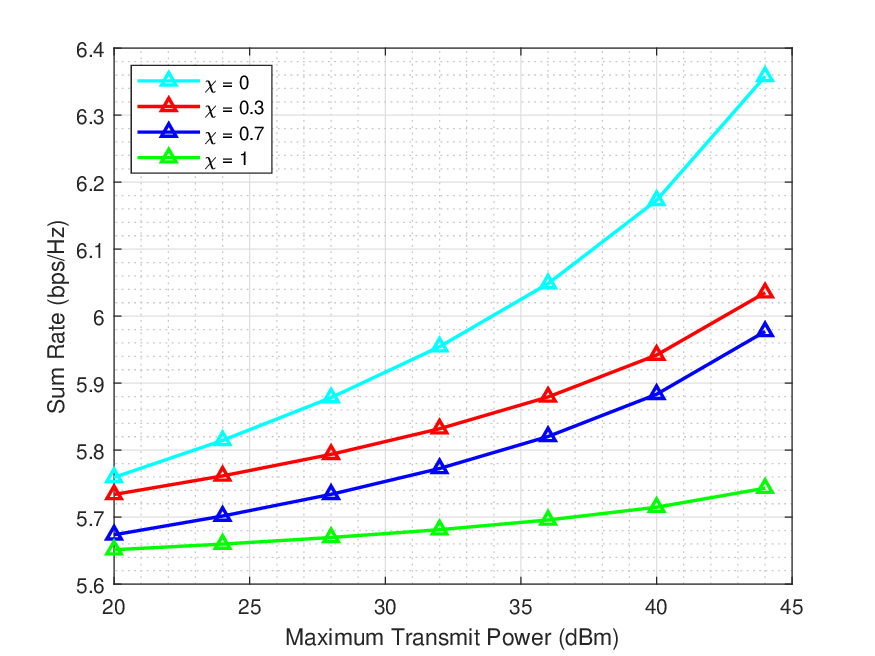} 
	\caption{RMSE versus maximum transmit power budget with different communication-sensing trade-off factor.}
	\label{tradeoff}
\end{figure}

To coordinate the mutual influence between communication and sensing, we define the trade-off factor as $\chi=\frac{\nu}{\sqrt{\nu^2+\varepsilon_1^2+\varepsilon_2^2}}$ with the $\nu$ being threshold of the probability of communication interruption for maritime users. {In practical applications, $\chi$ is dynamically adjusted to suit task-specific priorities. Specifically, for safety-critical tasks like navigation and route planning, a larger $\chi$ ensures high sensing accuracy. Whereas, for communication-intensive scenarios such as business data transmission, a smaller $\chi$ is preferred to maximize the overall throughput.} It worth noting that $\chi=0$ and $\chi=1$ denote the pure communication and the pure location sensing, respectively. As shown in Fig. $\ref{tradeoff}$, when the trade-off factor $\chi$ decreases, the sum rate of maritime users increases. Under the same transmit power, when the higher requirement of communication is needed, there are more power allocated to communication. Furthermore, with the larger transmit power, performance of communication and sensing are both enhanced. Thus, in the practical applications, we need to formulate appropriate trade-off factor between communication and sensing based on different demands and scenarios.

\begin{figure}[h]
	\centering
	\includegraphics[width=0.5\textwidth]{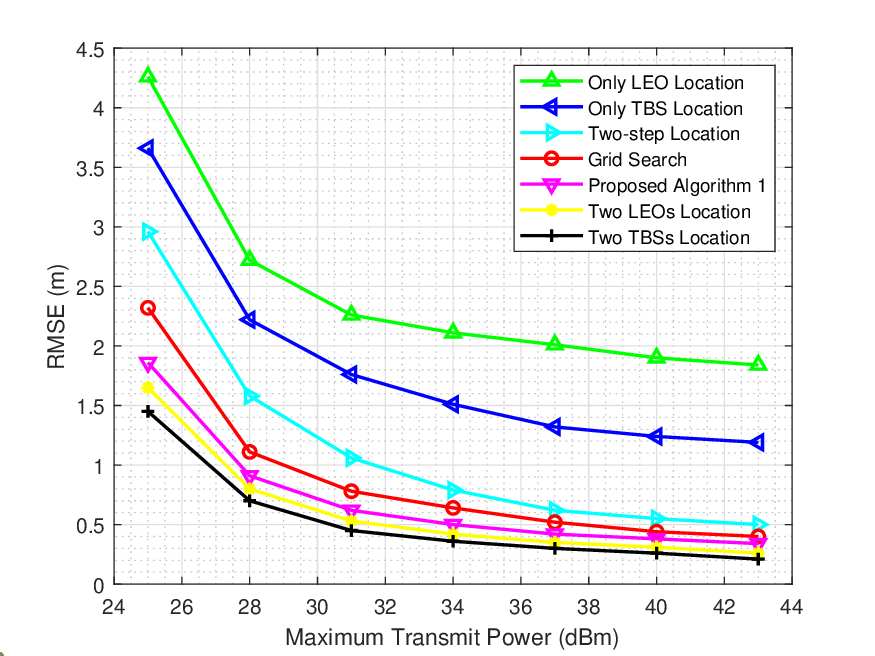} 
	\caption{RMSE versus maximum transmit power budget with different location sensing algorithms.}
	\label{duibi1xin}
\end{figure}

{In Fig. $\ref{duibi1xin}$, we focus on the RMSE obtained from the only LEO performing sensing, only TBS performing sensing, two LEOs performing sensing, two TBSs performing sensing, two-step location, grid search algorithm, and the proposed Algorithm 1.} Firstly, we can see that when relying only on a single base station for location sensing, the obtained RMSE of the target position is not accurate. It is because the location information of the target obtained by single base station from the echo signal is too little to locate the target coordinates. {Furthermore, it is observed that increasing the number of sensing nodes significantly enhances the localization accuracy compared to the single-node scenarios. This is because the multi-node collaborative sensing provides richer spatial diversity and more comprehensive geometric information from the reflected echo signals, which effectively reduces the localization uncertainty.} Although, the traditional two-step location can locate the target more accurately, the computing resources required have increased significantly. For the grid search method, the smaller search accuracy is set, the more accurate localization accuracy can be obtained. However, the required computing and time resources increase exponentially with small search accuracy. Thus, the proposed Algorithm 1 demonstrates higher localization accuracy in comparison to the other algorithms.

\begin{figure}[h]
	\centering
	\includegraphics[width=0.5\textwidth]{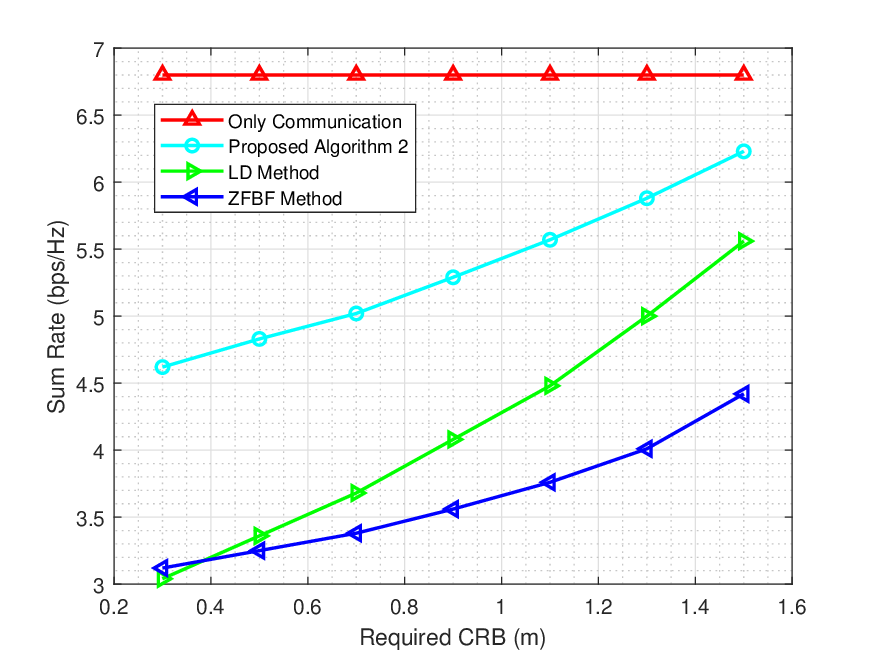} 
	\caption{performance of proposed Algorithm 2 versus required location accuracy with the baseline algorithms.}
	\label{duibi2}
\end{figure}

Finally, we evaluate Algorithm 2 with several existing beamforming techniques under identical simulation conditions. The LD approach \cite{LD} employs Lagrange duality to transform the beamforming design scenario into a more tractable optimization formulation. The zero-forcing beamforming (ZFBF) method is always used to design the transmit beamforming, which can completely eliminate the inter-beam interference among users and provide high sum rate performance \cite{ZFBF}. The result presented in Fig. $\ref{duibi2}$ indicates a consistent growth in the sum rate of maritime users for all algorithms when lower localization accuracy is required. Furthermore, the proposed Algorithm 2 has significant gains in terms of sum rate compared with the baseline algorithms, substantiating the effectiveness of the proposed Algorithm 2. Besides, the only communication scheme also demonstrates the upper bound of communication performance.

\section{Conclusion}

In this paper, we introduced an ISTMS with joint communication and sensing based on the same radio-frequency signals. According to the unique propagation environment at sea, we modeled the shore-to-ship channel with two-ray path-loss model. Based on the DE method, we proposed a sensing algorithm to obtain the delay and angle from echo signal. To evaluate the communication and sensing performance, we derived the closed-form expressions of the data transmission rate and CRB, respectively. Further, by jointly designing the beamforming at the TBS and the LEO satellite, we maximized the sum date transmission rate of maritime users while satisfying target localization accuracy requirements and transmit power constraints. Lastly, we presented lots of simulation results to demonstrate the effectiveness of proposed algorithms.

\begin{appendices}
	\section{Convergence Analysis of DE Method}
	We consider the minimization problem $\min_{\mathbf{x}\in\mathcal{X}} f(\mathbf{x})$, where $\mathcal{X}\subset\mathbb{R}^D$ is a bounded feasible region and $f$ is a real-valued function. The global minimum value is denoted as $f^* = \min_{\mathbf{x}\in\mathcal{X}} f(\mathbf{x})$, and the global optimum set as $\mathcal{X}^* = \{\mathbf{x}\in\mathcal{X}: f(\mathbf{x})=f^*\}$.
	
	\begin{algorithm}[!ht]
		\caption{DE Algorithm with Elitist Strategy}
		\label{alg:DE_improved}
		\begin{algorithmic}[1]
			\REQUIRE $N_p$, $F$, $CR \in (0,1)$, $G_{\max}$, $\mathcal{X}$.
			\ENSURE Best solution $\mathbf{x}_{\text{best}}$.
			\STATE Initialize $\{\mathbf{x}_i^0\}_{i=1}^{N_p}$ uniformly in $\mathcal{X}$, $g=0$.
			\WHILE{$g < G_{\max}$}
			\FOR{$i=1$ to $N_p$}
			\STATE Mutation: $\mathbf{v}_i^g = \mathbf{x}_{r_1}^g + F(\mathbf{x}_{r_2}^g - \mathbf{x}_{r_3}^g)$ with distinct $r_1,r_2,r_3\neq i$.
			\STATE Crossover: Generate $\mathbf{u}_i^g$ via binomial crossover with probability $CR$.
			\STATE Selection: $\mathbf{x}_i^{g+1} = \begin{cases} \mathbf{u}_i^g, & f(\mathbf{u}_i^g) \le f(\mathbf{x}_i^g) \\ \mathbf{x}_i^g, & \text{otherwise} \end{cases}$
			\ENDFOR
			\STATE Elitist: $\mathbf{x}_{\text{best}}^g = \arg\min_{\mathbf{x}\in P_{g+1}} f(\mathbf{x})$, overwrite a random individual with it.
			\STATE $g \leftarrow g+1$.
			\ENDWHILE
			\STATE \textbf{return} $\mathbf{x}_{\text{best}} = \arg\min_{\mathbf{x}\in P_g} f(\mathbf{x})$.
		\end{algorithmic}
	\end{algorithm}
	
	Define the state space of the population as $\mathcal{S} = \mathcal{X}^{N_p},$ which is finite in the mathematical sense. The population sequence $\{P_g\}_{g=0}^{\infty}$ forms a stochastic process. To prove the convergence of the DE algorithm, we first give out three lemmas.
	
	\emph{Lemma 1:} $\{P_g\}_{g=0}^{\infty}$ is a finite homogeneous Markov chain.
	
	\textbf{Proof:} The transition from \(P_g\) to \(P_{g+1}\) depends only on the current population \(P_g\) and the random operations with fixed parameters, it does not depend on earlier generations. The state space \(\mathcal{S}\) is finite because \(\mathcal{X}\) is bounded and closed, and in practice it is discretized by machine precision, but theoretically it remains countable and a finite Markov chain analysis applies.
	
	\emph{Lemma 2:} Let \(f_{\text{best}}^g = \min_{\mathbf{x}\in P_g} f(\mathbf{x})\). Then \(f_{\text{best}}^{g+1} \le f_{\text{best}}^g\) for all \(g\).
	
	\textbf{Proof:} The elitist strategy guarantees that \(\mathbf{x}_{\text{best}}^g \in P_{g+1}\), hence the minimum cannot increase.
	
	\emph{Lemma 3:} From any population state \(P_g\) that does \emph{not} contain a global optimum, there exists a constant probability \(p>0\), such that \(P_{g+1}\) contains a global optimum with probability at least \(p\).

	\textbf{Proof:} Because \(CR\in(0,1)\) and the mutation step can generate any point in \(\mathcal{X}\) with positive probability when \(F\neq 0\). Therefore, for any global optimum \(\mathbf{x}^*\in\mathcal{X}^*\), there is a positive probability that a trial vector \(\mathbf{u}_i^g\) equals \(\mathbf{x}^*\). The selection step will accept \(\mathbf{x}^*\) because \(f(\mathbf{x}^*) \le f(\mathbf{x}_i^g)\). The elitist strategy does not prevent this event. Since the state space is finite, we can take the minimum of these positive probabilities over all non-optimal states and denote it by \(p>0\). Hence, we have $\Pr(P_{g+1}\cap\mathcal{X}^* \neq \varnothing \mid P_g\cap\mathcal{X}^* = \varnothing) \ge p$.
	
	Based on the above lemmas, we prove the convergence of the DE algorithm with the elitist strategy, which can be mathematically expressed as
	\begin{equation}
		\Pr\left( \lim_{g\to\infty} f_{\text{best}}^g = f^* \right) = 1
	\end{equation}
	
	\textbf{Proof:} Define event $A_g = \{ f_{\text{best}}^g = f^* \}$, i.e., the population at generation $g$ already contains a global optimum. Due to the elitist strategy, once $A_g$ occurs, $A_k$ holds for all $k\ge g$. Thus, the event that convergence does \emph{not} occur is $\bigcap_{g=0}^{\infty} A_g^c$. Using Lemma 3, if $A_g$ has not occurred by generation $g$, then
	\begin{equation}
		\Pr(A_{g+1}^c \mid A_g^c) \le 1-p,
	\end{equation}
	because with probability at least $p$ a global optimum appears in the next generation. By induction,
	\begin{equation}
		\Pr\left( \bigcap_{k=0}^{g} A_k^c \right) \le (1-p)^{g+1}.
	\end{equation}
	Taking the limit $g\to\infty$ yields $ \Pr\left( \bigcap_{g=0}^{\infty} A_g^c \right) = 0.$ Therefore $\Pr\left( \bigcup_{g=0}^{\infty} A_g \right) = 1$, i.e., with probability 1 a generation containing a global optimum is reached in finite time. From that point onward the best value stays at $f^*$. Consequently, we obtain
	\begin{equation}
		\lim_{g\to\infty} f_{\text{best}}^g = f^*.
	\end{equation}
	The proof is completed.
\end{appendices}

\end{document}